\documentclass[amsthm]{elsart}
\usepackage{amssymb,amsfonts,amsmath,times}
\usepackage{color}
\theoremstyle{plain}

\newtheorem{theorem}{Theorem}[section]
\newtheorem{corollary}[theorem]{Corollary}
\newtheorem{lemma}[theorem]{Lemma}

\newtheorem{proposition}[theorem]{Proposition}
\theoremstyle{definition}\theoremstyle{assumption}

\newtheorem{definition}[theorem]{Definition}
\theoremstyle{remark}
\newtheorem{remark}[theorem]{Remark}
\numberwithin{equation}{section}

\begin{document}

\begin{frontmatter}

\title{\normalfont\LARGE
Gibbs Measures on Marked Configuration Spaces: Existence and
Uniqueness}
\runtitle{Gibbs Measures on Marked Configuration Spaces}

\author[Munchen]{Diana Conache}
\ead{diana.conache@tum.de}
\author[York]{Alexei Daletskii}
\ead{alex.daletskii@york.ac.uk}
\author[Bielefeld]{Yuri Kondratiev}
\ead{kondrat@math.uni-bielefeld.de}
\author[Bielefeld]{Tanja Pasurek}
\ead{tpasurek@math.uni-bielefeld.de}

\address[Munchen]{Zentrum Mathematik, Technische Universit\"{a}t M\"{u}nchen, D-85748 Garching, Germany}
\address[York]{Department of Mathematics, University of York, York YO10 5DD, UK}
\address[Bielefeld]{Fakult\"{a}t f\"{u}r Mathematik, Universit\"{a}t Bielefeld, D-33501 Bielefeld, Germany }

\thanks[t1]{Supported by the DFG through SFB 701 \textquotedblleft Spektrale
Strukturen und Topologische Methoden in der Mathematik\textquotedblright ,
IRTG 1132 \textquotedblleft Stochastics and Real World
Models\textquotedblright  (Universit\"{a}t Bielefeld).}

\runauthor{ D.\ Conache, A.\ Daletskii, Yu.\ Kondratiev, T.\Pasurek/Journal of Functional
Analysis}

\begin{abstract}
We study an infinite system of particles chaotically distributed over a
Euclidean space $\mathbb{R}^{d}$. Particles are characterized by their
positions $x\in \mathbb{R}^{d}$ and an internal parameter (spin) $\sigma
_{x}\in \mathbb{R}^{m}$, and interact via position-position and (position
dependent) spin-spin pair potentials. Equilibrium states of such system are
described by Gibbs measures on a marked configuration space.
Due to the presence of unbounded spins, the model does not fit the classical (super-) stability theory of Ruelle.
The main result of the paper is the derivation of sufficient conditions of the
existence and uniqueness of the corresponding Gibbs measures.
\end{abstract}

\begin{keyword}
\def\MSC{\par\leavevmode\hbox {\it 2000 MSC:\ }}%
Interacting particle systems, Gibbs measures, DLR equation,
Ruelle's superstability estimates, Dobrushin's uniqueness criterion,
random point processes, marked configuration spaces

\MSC Primary 82B44; Secondary 82B21, 82A57, 60G55
\end{keyword}

\end{frontmatter}

\section{Introduction}

The aim of this paper is to study the equilibrium states of the following
infinite particle system in continuum. We consider a countable collection $%
\gamma $ of identical point particles chaotically distributed over a
Euclidean space $X$ ($=\mathbb{R}^{d}$).$\ $Additionally, we assume that
each particle $x\in \gamma $ possesses an internal structure described by a
mark (spin) $\sigma _{x}$ taking values in a single-spin space $S$ ($=%
\mathbb{R}^{m}$) and characterized by a single-spin measure $\chi \ $on $S$.
Each two particles $x,y\in \gamma $ interact via a pair potential given by
the sum of two components:

(i) a purely positional (e.g. distance dependent, possibly singular or hard-core)  potential
\begin{equation}
\Phi :X\times X\rightarrow \mathbb{R}\cup \{+\infty \},\ \ \Phi (x,y)=\Phi
(y,x),\ \ x,y\in X  \label{inter-pos}
\end{equation}
(representing e.g. a molecular
force);

(ii) a (position-dependent) spin-spin interaction of the form $W_{xy}(\sigma
_{x},\sigma _{y}),$ where
\begin{equation}
W_{xy}=W_{yx}:S\times S\rightarrow \mathbb{R}\text{, \ \ }%
W_{xy}(s,t)=W_{xy}(t,s),\text{ \ }s,t\in S,  \label{inter-spin}
\end{equation}%
are symmetric functions of polynomial growth.

Our system can be seen as a combined type model, which carries features of
both an infinite particle system in continuum (i.e., non-ideal classical
gas) and an interacting system of unbounded spins on a discrete set (random
graph) formed by positions of the particles. Therefore we have to take into
account two possible catastrophic effects caused by dense particle
configurations and by strong spin interactions, respectively. Notably, our
model does not fit the setup of the previous papers on marked point
processes, which have mostly been dealing with the case of compact spins.
Thus its study requires development of new methods, involving an appropriate
concept of thermodynamical stability. The corresponding physical systems are
e.g. magnetic gases, ferrofluids, amorphous magnets, etc., see \cite{GH},
\cite{Gruber}, \cite{RZ}. Such compound (with additional spin variables)
models are of a special interest in mathematical physics because they
provide some (of still very few) examples of continuum systems where the
appearence of an (orientational odering) phase transition has been proved
rigorously. This makes important an alternative question of the absence of
phase transition, i.e. the uniqueness of thermal equilibrium states,
expected e.g. in the low density regime. Such models are still poorly
understood, to say nothing of the general case of non-compact (possibly
multi-dimentional vector) marks and unbounded (not necessarily ferromagnetic
or quadratic) spin interactions, which motivates our present study.

Once the interaction potentials have been specified, the whole system is
governed by the heuristic Hamiltonian%
\begin{equation*}
H(\widehat{\gamma }):=\sum_{\{x,y\}\subset \gamma }\Phi
(x,y)+\sum_{\{x,y\}\subset \gamma }W_{xy}(\sigma _{x},\sigma _{y})
\end{equation*}%
on the phase space $\widehat{\Gamma }(X)$ consisting of marked
configurations $\widehat{\gamma }=\{(x,\sigma _{x})\}$, where the
corresponding position configuration $\gamma =\left\{ x\right\} $ belongs to
the space%
\begin{equation*}
\Gamma (X):=\left\{ \gamma \subset X:\ N\left( \gamma _{\Lambda }\right)
<\infty \text{ for any }\Lambda \in \mathcal{B}_{0}(X)\right\} .
\end{equation*}%
Here $\mathcal{B}_{0}(X)$ is the collection of all compact subsets of $X$
and $N\left( \gamma _{\Lambda }\right) $ denotes the number of elements of $%
\gamma _{\Lambda }:=\gamma \cap \Lambda $. In what follows, we will use the
notation $\widehat{\gamma }_{\Lambda }:=\{(x,\sigma _{x}),\ x\in \gamma
_{\Lambda }\}.$

The equilibrium states of the system are described by certain probability
measures on $\widehat{\Gamma }(X)$. In absence of the interaction (the
so-called \textquotedblleft free\textquotedblright\ case), the equilibrium
state is unique and given by the marked Poisson measure%
\begin{equation*}
\widehat{\pi }(\mathrm{d}\widehat{\gamma })=\bigotimes_{x\in \gamma }\chi (%
\mathrm{d}\sigma _{x})\pi _{z}(\mathrm{d}\gamma ),
\end{equation*}%
where $\pi _{z}$ is the Poisson measure on $\Gamma (X)$ with intensity
(i.e., particle density) $z>0$, see e.g. \cite{DVJ1}, \cite{CKMS}. If the
interaction is present, the equilibrium states are given by marked Gibbs
measures $\mu $ on $\widehat{\Gamma }(X)$, which are constructed as
perturbations of $\widehat{\pi }$ by the (heuristic) density $\mathrm{exp}%
\left\{ -H(\widehat{\gamma })\right\} $. Rigorously, any such $\mu $ is a
probability measure on $\widehat{\Gamma }(X)$ with prescribed conditional
distributions $\mu (\mathrm{d}\widehat{\gamma }$ $|$ $\widehat{\gamma }=%
\widehat{\eta }$ off $\Lambda )$, $\widehat{\eta }\in \widehat{\Gamma }(X)$,
for an exhausting system of sets $\Lambda \in \mathcal{B}_{0}(X)$. These
conditional distributions, or Gibbs specification kernels of our model, are
explicitly given by formulae (\ref{specif}) and (\ref{specif0}) below and
will be denoted by $\Pi _{\Lambda }\left( \mathrm{d}\widehat{\gamma }%
\left\vert \widehat{\eta }\right. \right) $. So, the study of Gibbs measures
is reduced to the generic problem of reconstructing a Markov random field $%
\mu $ on $\widehat{\Gamma }(X)$ from its local specification $\Pi =\left\{
\Pi _{\Lambda }\right\} _{\Lambda \in \mathcal{B}_{0}(X)}$. This constitutes
the standard Dobrushin--Lanford--Ruelle (DLR) formalism described in details
in Section \ref{sec-gibbs}.

We denote by $\mathcal{G}$ the set of all such measures (for fixed $H$ and $%
\chi $). The study of the structure of the set $\mathcal{G}$ is of a great
importance. In particular, there are three fundamental questions arising
here:

\begin{itemize}
\item[(\textbf{E})] \textit{Existence}: is $\mathcal{G}$ not empty?

\item[(\textbf{U})] \textit{Uniqueness: }is\textit{\ }$\mathcal{G}$ a
singleton?

\item[(\textbf{M})] \textit{Multiplicity: }does\textit{\ }$\mathcal{G}$
contain at least two (and hence infinitely many) elements?
\end{itemize}

In this paper, we derive sufficient conditions for (E) and (U). We introduce
the set $\mathcal{G}^{\mathrm{t}}\subset \mathcal{G}$ of tempered Gibbs
measures that are concentrated on the space $\widehat{\Gamma }_{\mathrm{t}%
}(X)$ of configurations with certain bounds on their density and spin
growth, see (\ref{temper}), (\ref{temper1}). Under reasonable assumptions on
the interaction potentials $\Phi $ and $W$ (responsible for the global
stability of the system and listed under \textbf{(}A1\textbf{)}--\textbf{(}A6%
\textbf{)} below), we will prove that the set $\mathcal{G}^{\mathrm{t}}$ is
not empty (Theorem \ref{existence}) and, moreover, that $\mathcal{G}^{%
\mathrm{t}}$ is a singleton provided the particle density $z$ is small
enough (Theorem \ref{uniqueness}). To prove the existence, we use the
extension of the analytic method developed in \cite{KPR} for the case of
interacting particle systems without spins. A crucial technical step here is
to prove a uniform bound of certain exponential moments of the corresponding
specification kernels $\Pi _{\Lambda }\left( \mathrm{d}\widehat{\gamma }%
\left\vert \widehat{\eta }\right. \right) $ as $\Lambda \nearrow X$ for any
boundary condition $\widehat{\eta }\in \widehat{\Gamma }_{\mathrm{t}}(X)$.
This in turn allows to show the compactness (in the topology of local set
convergence on $\widehat{\Gamma }(X)$) of the family $\left\{ \Pi _{\Lambda
}\left( \mathrm{d}\widehat{\gamma }\left\vert \widehat{\eta }\right. \right)
,\ \Lambda \in \mathcal{B}_{0}(X)\right\} $ and thus the existence of the
limiting points, which can be identified with elements of $\mathcal{G}^{%
\mathrm{t}}$.

In order to study the uniqueness, we represent (via the natural embedding $%
\mathbb{Z}^{d}\subset X)$ the configuration space $\widehat{\Gamma }(X)$ in
the form $\widehat{\Gamma }(\mathrm{Q})^{\mathbb{Z}^{d}}$, where $\mathrm{Q}$
is an elementary cube in $X$, and construct a lattice model (with intricate
non-linear spin space $\widehat{\Gamma }(\mathrm{Q})$) equivalent to the
original continuum model. In this setting we can use the
Dobrushin--Pechersky approach to the uniqueness problem for lattice-type
systems, see \cite{DoP}, \cite[Theorem 2.6]{CKKP} and also \cite[Theorem 4]%
{PZ} and \cite[Theorem 3]{BP}, where this method is applied to continuum
systems (without spins) on $\Gamma (X)$. The uniform exponential moment
bounds allow us to control the interaction growth and to check the
conditions of the Dobrushin--Pechersky criterion for the lattice counterpart
of the continuum model. As a by-product of our method we also prove a decay
of correlations for the (unique) Gibbs measure (Corollary \ref{dec_corr}),
which seems to be entirely new for such systems.

Let us note that a general theory of Gibbs measures with the Ruelle-type
(super-) stable interactions on marked configuration spaces can be found
e.g. in \cite{AKLU}, \cite{KKdS}, \cite{KunaPhD} and \cite{Mase}. However,
it is essentially restricted to compact spins and hence does not apply to
our model (see Remark \ref{rem_bdd_spins}). The case of unbounded vector
spins interacting via potentials of superquadratic growth, including the
existence and uniqueness problems for the associated Gibbs states, has not
been treated so far in the literature.

Question (M) is discussed (for scalar spins and ferromagnetic interactions)
in a complementary paper \cite{DKK}.

The structure of the paper is as follows. In Section \ref{sec-model} we give
a rigorous description of our model (Subsections \ref{sec-conf}, \ref%
{sec-gibbs}, \ref{notations}) and formulate the main results (Subsection \ref%
{sec-results}). Section \ref{ME} is devoted to the derivation of moment
bounds. In Section \ref{sec-exist}, we prove our main result on the
existence problem -- Theorem \ref{existence}. Section \ref{sec-uniq} deals
with the uniqueness problem. We start with the lattice representation of our
model (Subsection \ref{sec-lattice}) and prove Theorem \ref{uniqueness} in
Subsection \ref{proof-uniq}. In Section \ref{sec-proofs}, we present proofs
of several technical lemmas.
\newpage

\section{The model and main results\label{sec-model}}

\subsection{Marked configuration spaces\label{sec-conf}}

As a location (phase) space $X$ for our particle system, let us fix the $d$%
-dimensional ($d\geq 1$) Euclidean space $\mathbb{R}^{d}$. It is endowed
with the Lebesgue measure $\mathrm{d}x$ on the Borel $\sigma $-algebra $%
\mathcal{B}(X)$. By $\mathcal{B}_{0}(X)$ we denote the ring of all bounded
sets from $\mathcal{B}(X)$. The configuration space $\Gamma (X)$ consists of
all locally finite subsets of $X$, that is,
\begin{equation}
\Gamma (X)=\left\{ \gamma \subset X:\ N\left( \gamma _{\Lambda }\right)
<\infty \text{ for any }\Lambda \in \mathcal{B}_{0}(X)\right\} ,
\label{gamma}
\end{equation}%
where $N\left( \gamma _{\Lambda }\right) $ stands for the cardinality of the
restriction $\gamma _{\Lambda }:=\gamma \cap \Lambda $. Let $C_{0}(X)$ be
the set of all continuous functions $f:X\rightarrow \mathbb{R}$ with compact
support. The space $\Gamma (X)$ is equipped in the standard way with the
\emph{vague topology}, which is the weakest one that makes continuous all
maps
\begin{equation*}
\Gamma (X)\ni \gamma \mapsto \left\langle f,\gamma \right\rangle
:=\sum_{x\in \gamma }f(x),\quad f\in C_{0}(X).
\end{equation*}%
It is well known (see, e.g., \cite[Section 15.7.7]{Kal}) that $\Gamma (X)$
is a Polish (i.e., separable completely metrizable) space in this topology;
an explicit construction of the appropriate metric can be found in \cite%
{KKut}. By $\mathcal{P}(\Gamma (X))$ we denote the space of all probability
measures on the corresponding Borel $\sigma $-algebra $\mathcal{B}(\Gamma
(X))$.

Let now $S$ be another Euclidean space $\mathbb{R}^{m}$ (with $m\neq d$ in
general) and consider the Cartesian product $\widehat{X}:=X\times S$. For
any element $\widehat{x}:=(x,s)$ of $\widehat{X}$ its $S$-component $s$ may
be seen as a mark (spin, charge etc.) attached to a particle placed at
position $x\in X$. Given a set $\Lambda \subset X$, we will often write for
short $\widehat{\Lambda }:=\Lambda \times S$. The canonical projection $%
p_{X}:X\times S\rightarrow X$ can be naturally extended to the configuration
space $\Gamma (\widehat{X}):=\Gamma (X\times S)$. Observe that for a
configuration $\widehat{\gamma }\in \Gamma (\widehat{X})$ its image $p_{X}(%
\widehat{\gamma })$ is a subset of $X$ that possibly admits accumulation and
multiple points, and hence does not in general belong to $\Gamma (X)$. The%
\emph{\ marked} configuration space\emph{\ }$\widehat{\Gamma }(X)$ is then
defined in the following way (see e.g. \cite{CKMS}, \cite{DVJ1}, \cite{KMM}):%
\begin{equation}
\widehat{\Gamma }:=\widehat{\Gamma }(X):=\left\{ \widehat{\gamma }\in \Gamma
(\widehat{X}):\text{ }p_{X}(\widehat{\gamma })\in \Gamma (X)\right\} .
\label{def-marked}
\end{equation}%
We will systematically use the notation%
\begin{equation*}
\gamma _{\Lambda }:=\gamma \cap \Lambda \text{ and }\widehat{\gamma }%
_{\Lambda }:=\widehat{\gamma }\cap \widehat{\Lambda }
\end{equation*}%
for $\gamma \in \Gamma (X),\ \widehat{\gamma }\in \widehat{\Gamma }(X),\
\Lambda \subset X$ and cylinder sets $\widehat{\Lambda }:=\Lambda \times S$%
.

We equip $\widehat{\Gamma }(X)$ with the so-called $\tau $-topology defined
as the weakest one that makes continuous the map%
\begin{equation}
\widehat{\Gamma }(X)\ni \widehat{\gamma }\mapsto \left\langle g,\widehat{%
\gamma }\right\rangle :=\sum_{(x,s)\in \widehat{\gamma }}g(x,s)  \label{top}
\end{equation}%
for any bounded continuous function $g:X\times S\rightarrow \mathbb{R}$ with
$\mathrm{supp\,}g\subset \Lambda \times S$ for some $\Lambda \in \mathcal{B}%
_{0}(X)$, i.e. with spatially compact support. This topology has been
employed in different frameworks in e.g.\textbf{\ }\cite{AKLU}, \cite{DKKP2}
and\ \cite{KunaPhD}; for a short account of its properties see also \cite%
{DKKP1}. An advantage of the $\tau $-topology is that it makes $\widehat{%
\Gamma }(X)$ a Polish space, in contrast to the vague topology inherited
from $\Gamma (\widehat{X})$ (which is generated by the maps (\ref{top}) with
$g\in C_{0}(\widehat{X})$). For an example of the $\tau $-consistent metric
on $\widehat{\Gamma }(X)$ see Section 2 of \cite{CG}. We then endow $%
\widehat{\Gamma }(X)$ with the associated Borel $\sigma $-algebra $\mathcal{B%
}(\widehat{\Gamma })$, also coinciding with the trace $\sigma $-algebra $%
\mathcal{B}(\Gamma (\widehat{X}))\cap \widehat{\Gamma }(X)$. This is the
smallest $\sigma $-algebra for which the counting variable
\begin{equation}
\widehat{\gamma }\mapsto N(\widehat{\gamma }\cap \Delta )  \label{ng}
\end{equation}%
is measurable for any $\Delta \in \mathcal{B}(X\times S)$ with $p_{X}(\Delta
)\in \mathcal{B}_{0}(X)$.

For a fixed $\Lambda \in \mathcal{B}_{0}(X)$, we consider the space%
\begin{equation}
\widehat{\Gamma }_{\Lambda }:=\widehat{\Gamma }_{\Lambda }(X)=\left\{
\widehat{\gamma }\in \widehat{\Gamma }(X):\text{ }p_{X}(\widehat{\gamma }%
)\subset \Lambda \right\}  \label{lo1}
\end{equation}%
of marked configurations located in the cylinder set $\widehat{\Lambda }%
:=\Lambda \times S$. It will be equipped with the image topology $p_{\Lambda
}\circ \tau $ induced from $\widehat{\Gamma }(X)$ under the natural
projection%
\begin{equation}
p_{\Lambda }:\ \widehat{\Gamma }(X)\ni \widehat{\gamma }\mapsto \widehat{%
\gamma }_{\Lambda }\in \widehat{\Gamma }_{\Lambda }(X)  \label{pl}
\end{equation}%
and with the corresponding $\sigma $-algebra $\mathcal{B}(\widehat{\Gamma }%
_{\Lambda })=\mathcal{B}(\widehat{\Gamma })\cap \widehat{\Gamma }_{\Lambda
}(X)$. Notably, $(\widehat{\Gamma }_{\Lambda }(X)$, $\mathcal{B}(\widehat{%
\Gamma }_{\Lambda }))$ is a \emph{standard Borel} space, which means that $%
\mathcal{B}(\widehat{\Gamma }_{\Lambda })$ can be generated by some
separable and complete metric on $\widehat{\Gamma }_{\Lambda }(X)$. We can
now define the $\sigma $-algebra $\mathcal{B}_{\Lambda }(\widehat{\Gamma }%
):=p_{\Lambda }^{-1}\circ \mathcal{B}(\widehat{\Gamma }_{\Lambda })$ on $%
\widehat{\Gamma }(X)$, which is constituted by the sets
\begin{equation}
\left\{ \widehat{\gamma }\in \widehat{\Gamma }(X):\widehat{\gamma }_{\Lambda
}\in \Delta \right\} ,\text{\quad }\Delta \in \widehat{\Gamma }_{\Lambda
}(X),  \label{pl1}
\end{equation}%
and hence is $\sigma $-isomorphic to $\mathcal{B}(\widehat{\Gamma }_{\Lambda
}(X))$. In other words, $\mathcal{B}_{\Lambda }(\widehat{\Gamma })\subset
\mathcal{B}(\widehat{\Gamma })$ is the smallest $\sigma $-algebra generated
by all variables (\ref{ng}) with $p_{X}(\Delta )\subset \Lambda .$ Then $(%
\widehat{\Gamma }(X),\mathcal{B}(\widehat{\Gamma }))$ can be seen as a
projective limit of the measurable spaces $(\widehat{\Gamma }_{\Lambda }(X),%
\mathcal{B}(\widehat{\Gamma }_{\Lambda }))$, $\Lambda \in \mathcal{B}_{0}(X)$%
, with respect to projection maps, cf. (\ref{pl}),
\begin{equation}
p_{\Lambda ^{\prime },\Lambda }:\widehat{\ \Gamma }_{\Lambda }(X)\ni
\widehat{\gamma }_{\Lambda }\mapsto \widehat{\gamma }_{\Lambda ^{\prime
}}\in \widehat{\Gamma }_{\Lambda ^{\prime }}(X),\text{ \ \ }\Lambda ^{\prime
}\subset \Lambda .  \label{pl2}
\end{equation}%
In particular, this allows us to use a version of \emph{Kolmogorov's theorem}
(cf. \cite[Theorem V.3.2]{Par}), according to which any probability measure $%
\mu \in \mathcal{P}(\widehat{\Gamma })$ is uniquely determined by its
projections $\mu _{\Lambda }:=p_{\Lambda }^{\ast }\mu \in \mathcal{P}(%
\widehat{\Gamma }_{\Lambda }),$ $\Lambda \in \mathcal{B}_{0}(X)$. Here and
in what follows, we denote by $\mathcal{P}(\widehat{\Gamma })$ and $\mathcal{%
P}(\widehat{\Gamma }_{\Lambda })$ the spaces of probaility measures on $%
\mathcal{B}(\widehat{\Gamma })$ and $\mathcal{B}(\widehat{\Gamma }_{\Lambda
}),$ respectively.

We will also need the subset of marked configurations \emph{finite} in all
of $\widehat{X}$
\begin{equation}
\widehat{\Gamma }_{0}:=\widehat{\Gamma }_{0}(X):=\bigcup_{\Lambda \in
\mathcal{B}_{0}(X)}\widehat{\Gamma }_{\Lambda }(X)  \label{locg}
\end{equation}%
and the subalgebra of \emph{local} events in $\widehat{\Gamma }(X)$
\begin{equation}
\mathcal{B}_{0}(\widehat{\Gamma }):=\bigcup_{\Lambda \in \mathcal{B}_{0}(X)}%
\mathcal{B}_{\Lambda }(\widehat{\Gamma }).  \label{loc}
\end{equation}

\begin{remark}
\label{fibre}The space $\widehat{\Gamma }(X)$ has a fibre bundle-type
structure over $\Gamma (X)$, where the fibres $p_{X}^{-1}(\gamma )$ can be
identified with the product spaces
\begin{equation*}
S^{\gamma }=\prod\limits_{x\in \gamma }S_{x},\ \ \ S_{x}:=S.
\end{equation*}%
Thus each $\widehat{\gamma }\in \widehat{\Gamma }(X)$ can be represented by
the pair
\begin{equation*}
\widehat{\gamma }=(\gamma ,\sigma _{\gamma }),\text{ \ \ where }\gamma
=p_{X}(\widehat{\gamma })\in \Gamma (X),\text{ \ }\sigma _{\gamma }=(\sigma
_{x})_{x\in \gamma }\in S^{\gamma }.
\end{equation*}%
It follows directly from the definition of the corresponding topologies that
the map $p_{X}:\widehat{\Gamma }(X)\rightarrow \Gamma (X)$ is continuous.
Hence for any configuration $\gamma $ the space $S^{\gamma
}=p_{X}^{-1}(\gamma )$ can be considered as a Borel subset of $\widehat{%
\Gamma }(X)$.
\end{remark}

From now on we fix a \emph{single-spin distribution} $\chi \in \mathcal{P}%
(S) $ ($=:$ the space of probability measures on $S$) and constant $z>0$
called the \textit{intensity} or \emph{activity parameter}. Observe that
each measurable $f:\widehat{\Gamma }_{0}(X)\rightarrow \mathbb{R}$ can be
identified with a family of symmetric Borel functions $f_{n}:(X\times
S)^{n}\rightarrow \mathbb{R}$, $n\in \mathbb{N}_{,}$ such that
\begin{equation*}
f(\widehat{\gamma })=f_{n}((x_{1},\sigma _{1}),\dots ,(x_{n},\sigma _{n}))%
\text{ for }\widehat{\gamma }=\{(x_{1},\sigma _{1}),\dots ,(x_{n},\sigma
_{n})\}.
\end{equation*}%
The \emph{marked Lebesgue-Poisson measure} $\widehat{\lambda }_z$ is defined
on $(\widehat{\Gamma }_{0}(X),\mathcal{B}(\widehat{\Gamma }_{0}))$ by the
relation
\begin{eqnarray}
&&\int_{\widehat{\Gamma }_{0}}f(\widehat{\gamma })\ \widehat{\lambda }_z(%
\mathrm{d}\widehat{\gamma })=f(\emptyset )  \label{LPM} \\[0.08in]
&&+\sum_{n=1}^{\infty }\frac{z^{n}}{n!}\int_{\left( X\times S\right)
^{n}}f_{n}((x_{1},\sigma _{1}),\dots ,(x_{n},\sigma _{n}))\ \chi (\mathrm{d}%
\sigma _{1})\mathrm{d}x_{1}\cdots \chi (\mathrm{d}\sigma _{n})\mathrm{d}%
x_{n},  \notag
\end{eqnarray}%
which has to hold for all measurable $f:\widehat{\Gamma }_{0}(X)\rightarrow
\mathbb{R}_{+}$. For each $\Lambda \in \mathcal{B}_{0}(X)$ it is a finite
measure on $\widehat{\Gamma }_{\Lambda }$ with mass $\widehat{\lambda }_z(%
\widehat{\Gamma }_{\Lambda })=\exp \left\{ z\int_{\Lambda }dx\right\} $.
Likewise, the Lebesgue-Poisson measure $\lambda _{z}$ on $\left( \Gamma
_{0}(X),\mathcal{B}(\Gamma _{0})\right) $ is defined by
\begin{equation}
\int_{\Gamma _{0}}f(\gamma )\lambda _{z}(\mathrm{d}\gamma )=f(\emptyset
)+\sum_{n=1}^{\infty }\frac{z^{n}}{n!}\int_{X^{n}}f_{n}(x_{1},\dots ,x_{n})\
\mathrm{d}x_{1}\cdots \mathrm{d}x_{n},  \label{LPMb}
\end{equation}%
holding for all measurable $f:\Gamma _{0}(X)\rightarrow \mathbb{R}_{+}$%
.

It is clear that $\lambda _{z}$ is an image of $\widehat{\lambda }_z$ under
the projection $p_{X}:\widehat{\Gamma }_{0}(X)\rightarrow \Gamma _{0}(X)$,
whereby $\widehat{\lambda }_z$ allows the disintegration
\begin{equation}
\widehat{\lambda }_z\left( \mathrm{d}\widehat{\gamma }\right)
:=\bigotimes\limits_{x\in \gamma }\chi (\mathrm{d}\sigma _{x})\ \lambda _{z}(%
\mathrm{d}\gamma ).  \label{LPMbb}
\end{equation}

\subsection{The model\label{sec-gibbs}}

Following the DLR\ approach (for its comprehensive exposition see \cite{Geor}%
), in this section we will give the rigorous definition of (grand canonical)
Gibbs measures associated with the interaction potentials (\ref{inter-pos}),
(\ref{inter-spin}) and a single-spin measure $\chi $.

We define the \emph{Hamiltonian} (or \emph{energy functional}) $H:\widehat{%
\Gamma }_{0}(X)\rightarrow \mathbb{R}$ by the formula%
\begin{equation}
H(\widehat{\gamma }):=U(\gamma )+E(\sigma _{\gamma }),\qquad \widehat{\gamma
}=(\gamma ,\sigma _{\gamma })\in \widehat{\Gamma }_{0}(X),
\label{big_ham_ann}
\end{equation}%
involving the positional and spin counterparts
\begin{equation}
U(\gamma ):=\sum_{\left\{ x,y\right\} \subset \gamma }\Phi (x,y)\text{ \ and
\ }E(\sigma _{\gamma }):=\sum_{\left\{ x,y\right\} \subset \gamma
}W_{xy}(\sigma _{x},\sigma _{y}),  \label{E_gamma_ann}
\end{equation}%
where the sums run over all (unordered) pairs of distinct points $x,y\in
\gamma $. By convention, we put $H(\{\emptyset \})=0$ and $H(\{(x,\sigma
_{x})\})=0$ for all $(x,\sigma _{x})\in \widehat{X}$.

For any $\Lambda \in \mathcal{B}_{0}(X)$ and $\widehat{\eta }=(\eta ,\xi
_{\eta })\in \widehat{\Gamma }(X)$, the relative local energy is given by%
\begin{equation}
H_{\Lambda }(\widehat{\gamma }_{\Lambda }|\widehat{\eta })=H(\widehat{\gamma
}_{\Lambda })+\mathrm{\Delta }H_{\Lambda }(\widehat{\gamma }_{\Lambda }|%
\widehat{\eta })  \label{HL1}
\end{equation}%
where%
\begin{equation}
\mathrm{\Delta }H_{\Lambda }(\widehat{\gamma }_{\Lambda }|\widehat{\eta }%
):=\sum_{x\in \gamma _{\Lambda }}\sum_{y\in \eta _{\Lambda ^{\mathrm{c}%
}}}\Phi (x,y)+\sum_{x\in \gamma _{\Lambda }}\sum_{y\in \eta _{\Lambda ^{%
\mathrm{c}}}}W_{xy}(\sigma _{x},\xi _{y}).  \label{DH}
\end{equation}%
Separating different types of interactions, we may rewrite (\ref{HL1}) as
\begin{equation}
H_{\Lambda }(\widehat{\gamma }_{\Lambda }|\widehat{\eta })=U_{\Lambda
}(\gamma _{\Lambda }|\eta )+E_{\Lambda }(\sigma _{\Lambda }|\xi )
\label{HL2}
\end{equation}%
with%
\begin{eqnarray}
U_{\Lambda }(\gamma _{\Lambda }|\eta ) &=&U(\gamma _{\Lambda })+\sum_{x\in
\gamma _{\Lambda }}\sum_{y\in \eta _{\Lambda ^{\mathrm{c}}}}\Phi (x,y),
\label{U1} \\
E_{\Lambda }(\sigma _{\gamma _{\Lambda }}|\xi ) &=&E_{\Lambda }(\sigma
_{\gamma _{\Lambda }})+\sum_{x\in \gamma _{\Lambda }}\sum_{y\in \eta
_{\Lambda ^{\mathrm{c}}}}W_{xy}(\sigma _{x},\xi _{y}).  \label{E1}
\end{eqnarray}%

The \emph{local Gibbs state} $\mu _{\Lambda }^{\widehat{\eta }}\in \mathcal{P%
}(\widehat{\Gamma }_{\Lambda })$ with boundary condition $\widehat{\eta }\in
\widehat{\Gamma }(X)$ fixed outside volume $\Lambda \in \mathcal{B}_{0}(X)$
is defined by the formula
\begin{equation}
\mu _{\Lambda }^{\widehat{\eta }}\left( \mathrm{d}\widehat{\gamma }_{\Lambda
}\right) :=Z_{\Lambda }(\widehat{\eta })^{-1}\mathrm{exp~}\left\{
-H_{\Lambda }(\widehat{\gamma }_{\Lambda }|\widehat{\eta })\right\} \
\widehat{\lambda }_{z,\Lambda }(\mathrm{d}\widehat{\gamma }_{\Lambda }),
\label{loc-gibbs}
\end{equation}%
where $\widehat{\lambda }_{z,\Lambda }$ is the restriction of the
Lebesgue-Poisson measure $\widehat{\lambda }_z$ to $\mathcal{B}(\widehat{%
\Gamma }_{\Lambda })$. We will often omit the subscript $\Lambda $ and just
write $\widehat{\lambda }_z\left( \mathrm{d}\widehat{\gamma }_{\Lambda
}\right) $ and $\lambda (d\gamma _{\Lambda })$. Here%
\begin{equation}
Z_{\Lambda }(\widehat{\eta }):=\int_{\widehat{\Gamma }_{\Lambda }}\mathrm{exp%
}~\left\{ -H_{\Lambda }(\widehat{\gamma }_{\Lambda }|\widehat{\eta }%
)\right\} \ \widehat{\lambda }_z(\mathrm{d}\widehat{\gamma }_{\Lambda })
\label{norm_fact}
\end{equation}%
is the normalizing factor (called the \emph{partition function}) making $\mu
_{\Lambda }^{\widehat{\eta }}$ a probability measure on $\widehat{\Gamma }%
_{\Lambda }(X)$ (provided $Z_{\Lambda }(\widehat{\eta })<\infty $, which
will be the case under certain conditions on the interaction potentials, cf.
Corollary \ref{part}). Next, we introduce stochastic kernels
\begin{equation*}
\widehat{\Gamma }(X)\times \mathcal{B(}\widehat{\Gamma })\ni (\widehat{\eta }%
,B)\mapsto \Pi _{\Lambda }\left( B|\widehat{\eta }\right) \in \lbrack 0,1]
\end{equation*}%
by the formula%
\begin{equation}
\Pi _{\Lambda }\left( B|\widehat{\eta }\right) :=\mu _{\Lambda }^{\hat{\eta}%
}\left( B_{\Lambda ,\widehat{\eta }}\right) ,\ \text{\ }B\in \mathcal{B(}%
\widehat{\Gamma }),  \label{specif}
\end{equation}%
where $B_{\Lambda ,\widehat{\eta }}:=\left\{ \widehat{\gamma }_{\Lambda }:%
\widehat{\gamma }_{\Lambda }\cup \widehat{\eta }_{\Lambda ^{\mathrm{c}}}\in
B\right\} \in \mathcal{B(}\widehat{\Gamma }_{\Lambda })$. By construction,
the projection of $\Pi _{\Lambda }\left( \cdot |\widehat{\eta }\right) $ on $%
\widehat{\Gamma }_{\Lambda ^{\mathrm{c}}}$ is just the $\delta $-measure
concentrated at $\widehat{\eta }_{\Lambda ^{\mathrm{c}}}$. So, the integral
relation
\begin{multline}
\int_{\widehat{\Gamma }}F(\widehat{\gamma })\Pi _{\Lambda }\left( \mathrm{d}%
\widehat{\gamma }|\widehat{\eta }\right)   \label{specif0} \\
=Z_{\Lambda }(\widehat{\eta })^{-1}\int_{\widehat{\Gamma }_{\Lambda }}F(%
\widehat{\gamma }_{\Lambda }\cup \widehat{\eta }_{\Lambda ^{\mathrm{c}}})%
\mathrm{exp}~\left\{ -H_{\Delta }(\widehat{\gamma }_{\Lambda }|\widehat{\eta
})\right\} \ \widehat{\lambda }_z\left( \mathrm{d}\widehat{\gamma }_{\Lambda
}\right) ,
\end{multline}%
holds for any measurable function $F:\widehat{\Gamma }(X)\rightarrow \mathbb{%
R}_{+}$. Furthermore, the map $\widehat{\Gamma }(X)\ni \hat{\eta}\mapsto \Pi
_{\Lambda }\left( B|\hat{\eta}\right) $ is measurable for each fixed $B\in
\mathcal{B}(\widehat{\Gamma })$.

The family $\Pi =\left\{ \Pi _{\Lambda }\right\} _{\Lambda \in \mathcal{B}%
_{0}(X)}$ constitutes a \emph{Gibbsian specification} on $\widehat{\Gamma }%
(X)$ (in the standard sense of \cite{Geor}, \cite{Pre}). In particular, it
obeys the \emph{consistency} property
\begin{equation}
\int_{\widehat{\Gamma }}\Pi _{\Lambda }\left( B|\widehat{\gamma }\right)
\ \Pi _{\Lambda ^{\prime }}\left( \mathrm{d}\widehat{\gamma }|\widehat{\eta }%
\right) =\Pi _{\Lambda^{\prime } }\left( B|\widehat{\eta }\right) ,  \label{consist}
\end{equation}%
which holds for any $B\in \mathcal{B}(\widehat{\Gamma })$, $\widehat{\eta }%
\in \widehat{\Gamma }(X)$ and $\Lambda ,\Lambda ^{\prime }\in \mathcal{B}%
_{0}(X)$ such that $\Lambda \subset \Lambda ^{\prime }$ (and thus $\widehat{%
\Lambda }\subset \widehat{\Lambda ^{\prime }}$).

Let $\mu $ be a probability measure on $\widehat{\Gamma }(X)$. We say that $%
\mu $ is a \emph{Gibbs state} associated with the specification $\Pi $ if it
satisfies the \emph{Dobrushin--Lanford--Ruelle} (DLR) equation
\begin{equation}
\mu (B)=\int_{\widehat{\Gamma }}\Pi _{\Lambda }\left( B|\widehat{\gamma }%
\right) \ \mu (\mathrm{d}\widehat{\gamma })  \label{DLR}
\end{equation}%
for all $B\in \mathcal{B}(\widehat{\Gamma })$ and $\Lambda \in \mathcal{B}%
_{0}(X)$. We denote by $\mathcal{G}:=\mathcal{G}(\widehat{\Gamma })$ the set
of all such measures.

In the \textquotedblleft free\textquotedblright\ case when both $\Phi $ and $%
W$ vanish, the corresponding unique Gibbs state $\mu \in \mathcal{G}$ is
just the \emph{marked Poisson measure} $\widehat{\pi }$. Equation (\ref{DLR}%
) then simplifies to Kolmogorov's theorem, which says that $\widehat{\pi }$
is fully determined by its local projections $\widehat{\pi }_{\Lambda }=[%
\widehat{\lambda }_z(\widehat{\Gamma }_{\Lambda })]^{-1}\widehat{\lambda }%
_{z,\Lambda }\in \mathcal{P(}\widehat{\Gamma }_{\Lambda }),$ $\Lambda \in
\mathcal{B}_{0}(X)$.

\subsection{Assumptions on the interaction \label{sec-assumptions}}

Let us specify conditions on the interaction potentials $\Phi ,W$ and
single-spin distribution $\chi $ to be used in the proof of our main
results. For that, we define a partition $\left( \mathrm{Q}_{k}\right)
_{k\in \mathbb{Z}^{d}}$ of $X$ by "elementary" volumes. Here $\mathrm{Q}_{k}$
is the half-open cube in $X$ with side length $1$ centered at point $%
k=(k^{(1)},...,k^{(d)})\in \mathbb{Z}^{d}\subset X$, that is,
\begin{equation}
\mathrm{Q}_{k}:=\left\{ x=(x^{(1)},...,x^{(d)})\in X:\ x^{(i)}\in \left[
k^{(i)}-1/2,\text{\thinspace }k^{(i)}+1/2\right) \right\} .  \label{cube}
\end{equation}%
For $k\in \mathbb{Z}^{d}$ and $\gamma \in \Gamma (X)$ resp. $\widehat{\gamma
}\in \widehat{\Gamma }(X)$, we then write for short
\begin{equation*}
\gamma _{k}:=\gamma _{\mathrm{Q}_{k}}\in \Gamma (\mathrm{Q}_{k})=:\Gamma _{k}%
\text{ \ resp. \ }\widehat{\gamma }_{k}:=\widehat{\gamma }_{\mathrm{Q}%
_{k}}\in \widehat{\Gamma }_{\mathrm{Q}_{k}}(X)=:\widehat{\Gamma }_{k}.
\end{equation*}%

In what follows we always assume that the following conditions hold.

\begin{enumerate}
\item[\textbf{(A1)}] \textit{Finite range of interactions}, that is, $%
\exists \ R>0$ such that $\Phi (x,y)=0$ and $W_{xy}=0$ if $\left\vert
x-y\right\vert \geq R$.

\item[\textbf{(A2)}] \textit{Lower boundedness} of $\Phi $, that is, $%
\exists \ M\geq 0$ such that
\begin{equation}
\inf_{x,y\in X}\Phi (x,y)\geq -M.  \label{M}
\end{equation}

\item[\textbf{(A3)}] \textit{Local strong superstability} of $U$, that is, $%
\mathcal{\exists }$ $P>2\ $such that for some $A_{\Phi }>0$ and $B_{\Phi
}\geq 0$
\begin{equation}
U(\gamma _{k})\geq A_{\Phi }N(\gamma _{k})^{P}-B_{\Phi }N(\gamma _{k})
\label{LSSS}
\end{equation}%
for any $k\in \mathbb{Z}^{d}$ and $\gamma \in \Gamma (X)$.

\item[\textbf{(A4)}] \textit{Uniform} \textit{polynomial bound} on $%
W_{xy}^{-}:=-\min \left\{ W_{xy},\ 0\right\} $, that is, $\exists $ $r>0$
and $\mathcal{J}$, $C_{W}\geq 0$ such that
\begin{equation}
W_{xy}^{-}(s,t)\leq \mathcal{J}\left( \left\vert s\right\vert
^{r}+\left\vert t\right\vert ^{r}+C_{W}\right) ,\ \ s,t\in S,  \label{A3}
\end{equation}%
for all $\{x,y\}\subset X$.

\item[\textbf{(A5)}] \textit{Exponential moment bound }on $\chi $, that is, $%
\exists ~q>r$ such that
\begin{equation}
\int_{S}e^{A_{\chi }\left\vert s\right\vert ^{q}}\chi (\mathrm{d}s)<\infty
\label{g-bound}
\end{equation}%
for some $A_{\chi }>0$.
\end{enumerate}

In addition, we require the following condition, which guarantees a \textit{%
spin-position superstability }type estimate (\ref{bound1}) crucial for our
method:

\begin{enumerate}
\item[\textbf{(A6)}] \label{A6}$P$,$\ q$ and $r$ satisfy the relation
\begin{equation}
(P-2)\left( q/r-1\right) >1.  \label{pq-rel}
\end{equation}%

\end{enumerate}

Let us point out that neither translation invariance nor continuity of $\Phi
$ and $W$ is assumed.

\begin{remark}
\label{super}(i) For every potential $\Phi $ obeying (A1) and (A2), the
local strong superstability (A3) readily implies the global one. More
precisely, for any $A_{\Phi }^{\prime }\in (0,A_{\Phi })$ there exists a $%
B_{\Phi }^{\prime }\geq 0$ such that%
\begin{equation}
U(\gamma )\geq A_{\Phi }^{\prime }\sum_{k\in \mathbb{Z}^{d}}N(\gamma
_{k})^{P}-B_{\Phi }^{\prime }N(\gamma ),\text{ \ \ }\gamma \in \Gamma
_{0}(X).  \label{sss}
\end{equation}%
This can be easily seen from the following chain of estimates
\begin{multline}
U(\gamma )\geq \sum_{k\in \mathbb{Z}^{d}}\left[ A_{\Phi }N(\gamma
_{k})^{P}-B_{\Phi }N(\gamma _{k})\right] -M\sum_{k\in \mathbb{Z}%
^{d}}\sum_{j\in \partial k}N(\gamma _{k})N(\gamma _{j})  \label{gsss} \\
\geq \sum_{k\in \mathbb{Z}^{d}}\left[ A_{\Phi }N(\gamma _{k})^{P}-M\mathcal{N%
}_{0}N(\gamma _{k})^{2}-B_{\Phi }N(\gamma _{k})\right]  \\
\geq (A_{\Phi }-\delta )\sum_{k\in \mathbb{Z}^{d}}N(\gamma _{k})^{P}-\left[
\left( M\mathcal{N}_{0}\delta ^{-1}\right) ^{\frac{2}{P-2}}+B_{\Phi }\right]
N(\gamma ),
\end{multline}%
where in the last line we used Young's inequality (\ref{tp2}) and $\mathcal{N}_{0}:=N(\partial k)$ is cardinality of the set $\partial k$. By choosing
small values of $\delta >0,$ we can get $A_{\Phi }^{\prime }$ arbitrarily
close to $A_{\Phi }.$\smallskip \newline
\quad (ii) The size of the elementary cubes in the partition $X=\coprod_{k\in
\mathbb{Z}^{d}}\mathrm{Q}_{k}$ is irrelevant. Fix any $\epsilon >0$, then
(A3) clearly holds for all $\mathrm{Q}_{k}^{\epsilon }:=\epsilon (\mathrm{Q}%
_{0}+k)$, $k\in \mathbb{Z}^{d}$, with proper constants $A_{\Phi ,\epsilon }>0
$ and $B_{\Phi ,\epsilon }\geq 0.$\smallskip \newline
\quad (iii) One of the best-understood examples of strong superstable
interactions is given by the so-called \emph{Dobrushin--Fisher--Ruelle}
(DFR) potentials behaving at the diagonal like $\Phi \left( x,y\right) \geq
c\left\vert x-y\right\vert ^{-d(1+\theta )}$ as $\left\vert x-y\right\vert
\rightarrow 0$, in which case $P=2+\theta $. For a detailed study and
historical comments see \cite{RT} and also \cite[Remark 4.1]{KPR}.\smallskip
\newline
\quad (iv) Assamption (A5) is aimed to compensate the polynomial growth of $W^-$ allowed by (A4). It is obvious that any measure satisfying condition (\ref%
{g-bound}) is finite. Thus without loss of generality we can choose $\chi $ to be a
probability measure. Furthermore, it is typically assumed that $\chi (%
\mathrm{d}s):=e^{-V(s)}\mathrm{d}s$ for some self-interaction potential $V:S%
\mathbb{\rightarrow R}$ growing fast enough:%
\begin{equation}
\exists ~A_{V}>A_{\chi }\text{ and }B_{V}\geq 0:\text{ \ }V(s)\geq
A_{V}\left\vert s\right\vert ^{q}-B_{V},\ \ s\in S.\text{ }  \label{V}
\end{equation}%
\smallskip \newline
\quad (v) The case of bounded $W_{xy}^{-}$ is essentially easier to handle. It can be
covered by a (simplified) version of our method, which will also work for $%
q=0,\ P=2$ (excluded from the general case by condition (\ref{pq-rel})).
This requires however $A_{\Phi }$ to be large enough. On the other hand,
this case fits into Ruelle's superstability approach extended in a
straightforward manner to marked configuration spaces (see a related comment
in Section \ref{sec-comments}) \smallskip \newline
\quad (vi) Except for the finite range, we impose no further restrictions on
the positive part $W_{xy}^{+}:=\max \left\{ W_{xy},\ 0\right\} $ of the
spin-spin interaction. Indeed, adding any $W_{xy}^{+}\geq 0$ could only
improve our basic estimates in Section \ref{sec-results}. Of a special
interest here are ferromagnetic interactions $W_{xy}$ of the form $%
J_{xy}|s-t|^{2}$ or $-J_{xy}\langle s,t\rangle $ with $J_{xy}\geq 0$
(notably, these two cases are not equivalent for our model insofar they
cannot be reduced to each other by changing the single-spin measure $\chi $%
), see also Remark \ref{rem2}. \smallskip \newline
\quad (vii) Assumption (\ref{pq-rel}) is crucial for our method. It excludes
the possibilty of $\Phi \equiv 0$ (that is, $P=0$, cf. (\ref{LSSS})), which
case can however be treated by modified arguments provided the spin-spin
interaction is purely repulsive, that is, $W_{xy}\geq 0$ (as pointed out in
Remark \ref{rem_4}). \smallskip \newline
\quad (viii) The case of multi-particle potentials $\Phi (x_{1},...x_{n})$
and $W(s_{1},...s_{n})$ with $n>2$ can be studied by similar methods
provided the superstability estimate of Proposition \ref{prop_main_est}
holds for the corresponding local Hamiltonians.\smallskip \newline
\quad (ix)
All the results below remain true if we take any non-atomic Radon measure $\sigma (\mathrm{d} x)$ on $(X,B(X))$ obeying the bound $sup_{k\in \mathbb{Z}^{d}} \,\sigma(Q_k)<\infty$ as intensity measure of the point process $\lambda_z$ (instead of the Lebesgue mass $dx$).
\end{remark}

\subsection{Notations\label{notations}}

Throughout the paper, we will use following shorthand notations (related to $%
\Lambda \in \mathcal{B}_{0}(X)$ and $k\in \mathbb{Z}^{d}$):\smallskip
\newline
$\ \ \ \Gamma :=\Gamma (X);$ $\widehat{\Gamma }:=\widehat{\Gamma }(X)\newline
$ $\ \ \Gamma _{\Lambda }:=\Gamma _{\Lambda }(X);$ $\gamma _{\Lambda
}:=\gamma \cap \Lambda \newline
$ $\ \ \widehat{\Gamma }_{\Lambda }:=\widehat{\Gamma }_{\Lambda }(X);$ $%
\widehat{\gamma }_{\Lambda }:=\widehat{\gamma }\cap (\Lambda \times S)%
\newline
$ $\ \ \Gamma _{k}:=\Gamma _{\mathrm{Q}_{k}};\ \gamma _{k}:=\gamma _{\mathrm{%
Q}_{k}}$\newline
$\ \ \ \widehat{\Gamma }_{k}:=\widehat{\Gamma }_{\mathrm{Q}_{k}};\ \widehat{%
\gamma }_{k}:=\widehat{\gamma }_{\mathrm{Q}_{k}}$\newline
$\ \ \ \partial k:=\left\{ j\neq k:\mathrm{dist}~(\mathrm{Q}_{k},\mathrm{Q}%
_{j})\leq R\right\} ,$ where `$\mathrm{dist}$' is the Euclidean distance
between two sets in $\mathbb{R}^{d}$\newline
$\ \ \mathcal{N}_{0}:=N(\partial k)$- cardinality of the set $\partial k$; obviously, it is independent of $%
k\in \mathbb{Z}^{d}\ $ and finite;\newline
$\ \ \ \gamma _{\partial k}:=\cup _{j\in \partial k}\gamma _{j}$; $\widehat{%
\gamma }_{\partial k}:=\cup _{j\in \partial k}\widehat{\gamma }_{j}$\newline
$\ \ \ \partial \Lambda :=\Lambda _{R}\setminus \Lambda =\Lambda _{R}\cap
\Lambda ^{\mathrm{c}}$\newline
\ \ $\ |\Lambda |:=\int_{\Lambda }\mathrm{d}x$ -- volume of $\Lambda \newline
$ $\ \ \mathrm{Q}_{\mathcal{K}}:=\bigcup_{j\in \mathcal{K}}\mathrm{Q}_{j},$ $%
\mathcal{K}\subset \mathbb{Z}^{d}\newline
$ $\ \ H_{k}(\widehat{\gamma }_{k}\left\vert \widehat{\eta }\right. ):=H_{%
\mathrm{Q}_{k}}(\widehat{\gamma }_{\mathrm{Q}_{k}}\left\vert \widehat{\eta }%
\right. )\newline
$ $\ \ U_{k}(\gamma _{k}\left\vert \eta \right. ):=U_{\mathrm{Q}_{k}}(\gamma
_{\mathrm{Q}_{k}}\left\vert \eta \right. ) $

Further notations will be introduced as needed.

\begin{remark}
\label{NR} By assumption (A1), both $\Phi (x,y)$ and $W_{xy}$ vanish for all
$x\in \mathrm{Q}_{k}$ and $y\in \mathrm{Q}_{j}$ whenever $j\notin \partial
k. $ The total number $\mathcal{N}_{0}=N(\partial k)$ of "neighbor" cubes $\mathrm{Q}_{j}$, $j\in \partial k$, %
 is independent of $k$ and can be roughly estimated by%
\begin{equation}
\mathcal{N}_{0}\leq v_{d}\left( R+\sqrt{d}\left/ 2\right.
\right) ^{d},\text{ \ \ }v_{d}=\frac{\pi ^{d/2}}{{\it{\Gamma }}\left(
1+d/2\right)},  \label{N0}
\end{equation}%
where $v_{d}$ is the volume of a unit ball in $\mathbb{R}^{d}$ and $\it{\Gamma }$ is the classical gamma function.
\end{remark}

\subsection{Main results\label{sec-results}}

Let us fix parameters $\kappa ,\vartheta >0$ and define \emph{control
functions} $F:\widehat{\Gamma }_{0}(X)\rightarrow \mathbb{R}_{+}$ and $%
F_{\alpha }:\widehat{\Gamma }(X)\rightarrow \mathbb{R}_{+}\cup \{+\infty \}$
by formulae
\begin{equation}
F(\widehat{\gamma })=\kappa N(\gamma )^{P}+\vartheta \sum_{x\in \gamma
}\left\vert \sigma _{x}\right\vert ^{q},\ \text{\ }\widehat{\gamma }=(\gamma
,\sigma ),  \label{Fk}
\end{equation}%
and
\begin{equation}
F_{\alpha }(\widehat{\gamma })=\underset{k\in \mathbb{Z}^{d}}{\mathrm{sup}}%
\left\{ e^{-\alpha \left\vert k\right\vert }F(\widehat{\gamma }_{k})\right\}
,\ \ \alpha >0,  \label{Fa}
\end{equation}%
respectively. Introduce the space of \emph{tempered configurations}%
\begin{equation}
\widehat{\Gamma }_{\mathrm{t}}(X):=\left\{ \widehat{\gamma }\in \widehat{%
\Gamma }(X):\ F_{\alpha }(\widehat{\gamma })<\infty \text{ for any }\alpha
>0\right\}  \label{temper}
\end{equation}%
and the corresponding set $\mathcal{G}^{\mathrm{t}}$ of \emph{tempered Gibbs
measures} that are supported by $\widehat{\Gamma }_{\mathrm{t}}(X)$, i.e.
\begin{equation}
\mathcal{G}^{\mathrm{t}}:=\left\{ \mu \in \mathcal{G}:\mu (\widehat{\Gamma }%
_{\mathrm{t}}(X))=1\right\} .  \label{temper1}
\end{equation}%
Obviously, the spaces $\widehat{\Gamma }_{\mathrm{t}}(X)$ and $\mathcal{G}^{%
\mathrm{t}}$ are independent of the choice of positive $\kappa $ and $%
\vartheta $. Furthermore, $\widehat{\Gamma }_{\mathrm{t}}(X)$ can be
characterized in the following way:%
\begin{equation}
\widehat{\Gamma }_{\mathrm{t}}(X)=\left\{ \widehat{\gamma }\in \widehat{%
\Gamma }(X):\ \sum_{k\in \mathbb{Z}^{d}}e^{-\alpha \left\vert k\right\vert
}F(\widehat{\gamma }_{k})<\infty \text{ for any }\alpha >0\right\} .
\label{Fb}
\end{equation}%

The next two theorems summarize the main results of this paper.

\begin{theorem}
\label{existence} (Existence and a priori estimate)

\begin{enumerate}
\item[(i)] The set $\mathcal{G}^{\mathrm{t}}$ is not empty.

\item[(ii)] For any given values%
\begin{equation}
\kappa \in (0,A_{\Phi })\text{ \ \ and \ \ }\vartheta \in (0,A_{\chi }),
\label{kt}
\end{equation}%
there exists a (explicitly computable) positive constant $\Psi :=\Psi
(\kappa ,\vartheta )$ such that each $\mu \in \mathcal{G}^{\mathrm{t}}$
obeys the moment estimate%
\begin{equation}
\underset{k\in \mathbb{Z}^{d}}{\mathrm{sup}}\int_{\widehat{\Gamma }}\exp
\left\{ F(\widehat{\gamma }_{k})\right\} \ \mu \left( \mathrm{d}\widehat{%
\gamma }\right) \leq \Psi .  \label{moment-est}
\end{equation}
\end{enumerate}
\end{theorem}

The proof will be given in Section \ref{sec-exist}. It is based on the
uniform bound of exponential moments for the corresponding specification
kernels (similar to (\ref{moment-est}), see Theorem \ref{theor1}) and local
equicontinuity of this specification (Theorem \ref{equicont2}), which in
turn implies that it possesses a cluster point $\mu \in \mathcal{G}^{\mathrm{%
t}}$.

\begin{theorem}
\label{uniqueness}(Uniqueness) For any given $\mathcal{J}_{0}>0$ there
exists $z_{0}=z_{0}(\mathcal{J}_{0})>0$ such that $\mathcal{G}^{\mathrm{t}}$
is a singleton for all $\mathcal{J}\leq \mathcal{J}_{0}$ and $z\leq z_{0}$.
\end{theorem}

\begin{remark}
\label{rem2} The threshold activity value $z_{0}$ can be computed
explicitly. Observe that $\left[ \lambda _{z}(\Gamma _{\Lambda })\right]
^{-1}\int_{\Gamma _{\Lambda }}N(\gamma _{\Lambda })\ \mathrm{d}\lambda
_{z}(\gamma _{\Lambda })=z$ for any $\Lambda \in \mathcal{B}_{0}(X)$, so
that $z$ can be interpreted as the point density of the underlying
Poisson point process, cf. \cite[p. 41]{CKMS}. Thus the uniqueness regime is
achieved in the systems with low particle density. On the other hand, for
large $z$ (that is, high particle density) one expects the existence of
multiple Gibbs states, see \cite{DKK} for the case of ferromagnetic
spin-spin interactions, where sufficient conditions of such multiplicity
(i.e., appearence of a phase transition) are given.
\end{remark}

Our proof of the uniqueness employs a lattice representation of our system and the
Dobrushin--Pecher\-sky criterion, see Section \ref{proof-uniq}. Sufficient
conditions of this criterion are checked using the moment bounds from
Section \ref{ME}.

\begin{remark}
\label{dec_corr}A result that seems to be completely new for this type of
systems is the decay of correlations of the Gibbs measures. Consider bounded
functions $G_{1},G_{2}:\widehat{\Gamma }(X)\rightarrow \mathbb{R}$, such
that $G_{1}$ is $\mathcal{B}_{\mathrm{Q}_{k_{1}}}(\widehat{\Gamma })$%
-measurable and $G_{2}$ is $\mathcal{B}_{\mathrm{Q}_{k_{2}}}(\widehat{\Gamma
})$-measurable, for some $k_{1},k_{2}\in \mathbb{Z}^{d}$. Let $\Vert \cdot
\Vert _{\infty }$ denote the usual $\mathrm{sup}$ norm. Set
\begin{equation*}
\mathrm{Cov}_{\mu }(G_{1};G_{2}):=\mu (G_{1}G_{2})-\mu (G_{1})\mu (G_{2})
\end{equation*}%
and assume that conditions of Theorem \ref{uniqueness} are satisfied. Let $%
\mu $ be the corresponding unique tempered Gibbs measure. Then, there exist
positive constants $\mathfrak{C}$ and $\mathfrak{a}$ such that
\begin{equation}
|\mathrm{Cov}_{\mu }(G_{1};G_{2})|\leq \mathfrak{C}\Vert G_{1}\Vert _{\infty
}\Vert G_{2}\Vert _{\infty }\exp \{-\mathfrak{a}|k_{1}-k_{2}|\}.  \label{cov}
\end{equation}%
This estimate is an immediate by-product of the (proof of) Theorem \ref%
{uniqueness} and follows from \cite[Theorem 2.7]{CKKP} adapted to our
setting via the lattice representation of the initial continuum model, see
Section \ref{sec-lattice}. Such approach (even in the case of a system
without marks) can be seen as a (simpler) alternative to the method of
clusters expansions (the only method by which similar results on $\Gamma (X)$
have been obtained).
\end{remark}

\subsection{Comments \label{sec-comments}}

\textbf{1.} \label{rem_bdd_spins} In \cite{AKLU, KKdS, KunaPhD, Mase}, a
theory of Gibbs measures (on marked configuration spaces) based on Ruelle's
classical approach (\cite{Ru69, Ru70}) has been elaborated. To this end, one
has to require either \emph{stability} or, moreover, \emph{superstability}
of the energy functional, expressed by the inequalities%
\begin{equation*}
H(\widehat{\gamma })\geq -C\cdot N(\gamma )
\end{equation*}%
\ and%
\begin{equation}
H(\widehat{\gamma })\geq A\sum_{k\in Z^{d}}N(\gamma _{k})^{2}-B\cdot
N(\gamma )  \label{super-stab}
\end{equation}%
respectively, holding for any $\widehat{\gamma }\in \widehat{\Gamma}_{0}(X)$
with some $A,B,C>0$. These bounds, which must be uniform in the variables $%
\sigma _{x}\in S$, obviously fail in the case of unbounded spin interactions
like in (\ref{big_ham_ann})--(\ref{E_gamma_ann}).

It seems to be possible to establish an analogue of Ruelle's superstability
estimates replacing the term $N(\gamma _{k})^{2}$ in (\ref{super-stab}) by
the control functional $F(\widehat{\gamma }_{k})$ (defined by (\ref{Fk}) and
involving both particles' positions and their spins). This will allow us to
construct the corresponding Gibbs states $\mu $ satisfying the regularity
condition
\begin{equation*}
\sup_{K\in \mathbb{N}}\left\{ K^{-d}\sum_{|k|\leq K}F(\widehat{\gamma }%
_{k})\right\} <\infty \ \ \text{for }\mu \text{-a.a. }\widehat{\gamma }\in
\widehat{\Gamma }(X).
\end{equation*}%
As for the uniqueness problem for such Gibbs states, one has to develop a
contraction theory of the Kirkwood--Salsburg equations for the corresponding
marked correlation functions. So far, this was only done in \cite{KunaPhD}
under condition (\ref{super-stab}) which, as already mentioned above, does
not cover our model.

\textbf{2.} Gibbs measures $\mu \in \mathcal{G}$ represent so-called \emph{%
annealed} thermodynamic states of our particle system; they describe the
thermal equilibrium of this system as a whole. Alternatively, one can
consider thermodynamic states of the spin system alone for a fixed typical
configuration (sample) $\gamma $, which is distributed according to a Gibbs
measure $\mu ^{\Phi }$ on $\Gamma (X)$ defined by the
position-position interaction $\Phi$. These are commonly referred to as
\emph{quenched} states, cf. \cite{BMR, Bov,New}. The corresponding Gibbs
measures $\mu _{\gamma }$ on the product spaces $S^{\gamma }$ were
constructed in \cite{DKKP1}. The relationship between Gibbs measures of
these two types can be expressed by the disintegration formula%
\begin{equation}
\mu \left( \mathrm{d}\widehat{\gamma }\right) =\mu _{\gamma }(\mathrm{d}%
\sigma _{\gamma })~{\cal M}(\mathrm{d}\gamma ),  \label{disf}
\end{equation}%
where ${\cal M}:=p_{X}^{\ast }\mu \in \mathcal{P}(\Gamma (X))$ is the projection of $%
\mu $ on $\Gamma (X)$, cf. Remark \ref{fibre} and \cite[formula (2.6)]{DKKP2}.
 In general, the projected measure ${\cal M}$ does not coincide with the Gibbs
measure $\mu ^{\Phi }$ and cannot be described in terms of position-position
interactions alone. Thus it is not clear whether the existence result from
\cite{DKKP1} could be used in order to prove the existence of the annealed
Gibbs measure $\mu $. Furthermore, (\ref{disf}) indicates that one cannot
directly compare (e.g., by means of various correlation inequalities known
for measures on $S^{\gamma }$, see e.g. \cite{Geor}, \cite{LP}) any two
annealed Gibbs states related to different spin-spin potentials $W_{xy}.$%

Let us remark that the multiplicity (phase transition) problem for quenched
Gibbs measures of ferromagnetic type has been studied in \cite{DKKP2}. On
the other hand, the question of uniqueness for quenched systems with
unbounded spins remains so far open. The main source of difficulties here
(making standard methods not applicable) is that the underlying discrete set
$\gamma \subset \mathbb{R}^{d}$ is highly inhomogeneous, so that $\mu ^{\Phi
}$-a.s. it holds $\sup_{k\in \mathbb{Z}^{d}}N(\gamma _{k})=+\infty $%
.

\textbf{3.} Analogously to the case of simple (i.e., unmarked) point
processes, one can show that each $\mu \in \mathcal{G}^{\mathrm{t}}$
satisfies the so-called \emph{Georgii--Nguen--Zessin} (GNZ) equation (see
e.g. \cite{KunaPhD, Mase}). It says that for any measurable function $G:%
\widehat{X}\times \widehat{\Gamma }\rightarrow \mathbb{R}_{+}$ the following
identity holds:%
\begin{multline}
\int_{\widehat{\Gamma }}\sum_{\widehat{x}\in \widehat{\gamma }}G(\widehat{x}%
,~\widehat{\gamma })~\mu \left( \mathrm{d}\widehat{\gamma }\right)
\label{GNZ} \\
=\int_{\widehat{\Gamma }}\int_{\widehat{X}}G(\widehat{x},~\widehat{\gamma }%
\cup \{\widehat{x}\})\exp \left\{ -\mathrm{\Delta }H(\{\widehat{x}\}|%
\widehat{\gamma })\right\} ~\mu \left( \mathrm{d}\widehat{\gamma }\right)
~\chi (\mathrm{d}\sigma _{x})\mathrm{d}x.  \notag
\end{multline}%
Here, cf. (\ref{DH}),%
\begin{equation*}
\mathrm{\Delta }H(\{\widehat{x}\}|\widehat{\gamma }):=\sum_{y\in \gamma }%
\left[ \Phi (x,y)+W_{xy}(\sigma _{x},\xi _{y})\right] ,\text{ \ \ }\widehat{%
\gamma }=(\eta ,\xi _{\gamma }).
\end{equation*}

\section{Exponential moment estimate\label{ME}}

\subsection{One-point estimates \label{sec-one-point}}

The following proposition is a starting point in the realization of our
approach. It describes the \emph{superstability} property of the system in
terms of the control functional $F$. The proof involves simple but tedious
calculations based on assumptions (A1)--(A6) and will be given in Section %
\ref{sec-proofs}.

\begin{proposition}
\label{prop_main_est} For any (arbitrarily small) $\delta >0$ one finds a
positive constant $C_{\delta }$ such that
\begin{equation}
-H_{k}(\widehat{\gamma }_{k}\left\vert \widehat{\eta }\right. )\leq -\left(
A_{\Phi }-\delta \right) N(\gamma _{k})^{P}  \label{bound1}
+\delta \sum_{x\in \gamma _{k}}\left\vert \sigma _{x}\right\vert ^{q}+\delta
\sum_{j\in \partial k}F(\widehat{\eta }_{j})+C_{\delta }
\end{equation}%
for all $k\in \mathbb{Z}^{d}$ and $\widehat{\gamma },\widehat{\eta }\in
\widehat{\Gamma }(X)$. Here $C_{\delta }:=C_{\delta }(\kappa ,\vartheta ;%
\mathcal{J})$ is a non-decreasing function of $\mathcal{J}$.
\end{proposition}

\begin{remark}
\label{rem1} Using the arguments from the proof of Proposition \ref%
{prop_main_est} (or, more precisely, Lemma \ref{superstab}) and the global
superstability of $U(\gamma )$ (see Remark \ref{super}), we get the bound%
\begin{multline}
-H_{\Lambda }(\widehat{\gamma }_{\Lambda }|\widehat{\eta })\leq -(A_{\Phi
}-\delta )\sum_{k\in \mathbb{Z}^{d}}N(\gamma _{\Lambda}\cap Q_k)^{P}+\delta
\sum_{x\in \gamma _{\Lambda }}\left\vert \sigma _{x}\right\vert
^{q}+C_{\Lambda ,\delta }(\widehat{\eta })  \label{HE-est} \\
\leq -(A_{\Phi }-\delta )P^{1-\mathcal{N}_{\Lambda }}N(\gamma _{\Lambda
})^{P}+\delta \sum_{x\in \gamma _{\Lambda }}\left\vert \sigma
_{x}\right\vert ^{q}+C_{\Lambda ,\delta }(\widehat{\eta }),
\end{multline}%
where $\mathcal{N}_{\Lambda }$  is the cardinality of the set 
$\left\{ j\in \mathbb{Z}^{d}:\mathrm{Q}_{j}\cap\Lambda )\ne \emptyset\right\} $. Both inequalities in (\ref{HE-est})
hold for an arbitrary domain $\Lambda \in \mathcal{B}_{0}(X)$, any $%
\widehat{\eta }\in \widehat{\Gamma }(X)$ and $\delta \in (0,A_{\Phi })$ with
an appropriate constant $C_{\Lambda ,\delta }(\widehat{\eta })\geq 0$ (the
explicit value of which is irrelevant for our purposes).
\end{remark}

Below we will frequently use the moment estimate
\begin{multline}
\int_{\widehat{\Gamma }_{\Lambda }}\exp \left\{ aN(\gamma _{\Lambda
})+b\sum_{x\in \gamma _{\Lambda }}\left\vert \sigma _{x}\right\vert
^{q}\right\} \ \widehat{\lambda }_z(\mathrm{d}\widehat{\gamma }_{\Lambda })
\label{finexp} \\
=\sum_{n=0}^{\infty }\frac{z^{n}}{n!}|\Lambda |^{n}e^{an}\left(
\int_{S}e^{b\left\vert s\right\vert ^{q}}\ \chi (\mathrm{d}s)\right)
^{n}=\exp \left\{ z|\Lambda |e^{a}\int_{S}e^{b\left\vert s\right\vert
^{q}}\ \chi (\mathrm{d}s)\right\} <\infty ,
\end{multline}%
which holds for any $\Lambda \in \mathcal{B}_{0}(X)$ and $a\in \mathbb{R}$, $%
b\leq A_{\chi }$ (cf. (\ref{g-bound})) and follows from the definition of
the Lebesgue-Poisson measure $\widehat{\lambda }_z$, assumption (A5) and
disintegration formula (\ref{LPMbb}).

\begin{corollary}
\label{part} The partition function $Z_{\Lambda }(\widehat{\eta })$
satisfies the estimate
\begin{equation}
1\leq Z_{\Lambda }(\widehat{\eta })<\infty  \label{part-bound}
\end{equation}%
for all $\Lambda \in \mathcal{B}_{0}(X)$ and $\widehat{\eta }\in \widehat{%
\Gamma }(X)$.
\end{corollary}

\textbf{Proof.} The lower bound can be immediately seen from the equalities $%
\widehat{\lambda }_{z,\Lambda }(\emptyset )=1$ and $U_{\Lambda }(\gamma
_{\Lambda }\left\vert \eta \right. )=E_{\gamma _{\Lambda }\cup \eta
_{\Lambda ^{c}}}(\sigma _{\gamma _{\Lambda }}\left\vert \xi \right. )=0$ if $%
\gamma _{\Lambda }=\emptyset $. The upper bound follows from (\ref{HE-est})
and (\ref{finexp}).
\hfill%
$\square  $

Lemmas \ref{DP1} and \ref{DP2} below provide us with crucial estimates on
the \textit{\textquotedblleft one-point\textquotedblright } kernels $\Pi
_{k}(\mathrm{d}\widehat{\gamma }|\widehat{\eta }):=\Pi _{\mathrm{Q}_{k}}(%
\mathrm{d}\widehat{\gamma }|\widehat{\eta }),$ $k\in \mathbb{Z}^{d},$
subject to varying boundary conditions $\widehat{\eta }\in \widehat{\Gamma }%
(X)$. To this end, let us fix some $\kappa \in (0,A_{\Phi })$ and $\vartheta
\in (0,A_{\chi })$ in definition (\ref{Fk}) of the functional $F$, cf. (\ref%
{kt}).$ $

\begin{lemma}
\label{DP1} For any (arbitrarily small) $\delta >0$ there exists a constant $%
\Xi _{\delta }>0$ such that for all $k\in \mathbb{Z}^{d}$ and $\widehat{\eta
}\in \widehat{\Gamma }(X)$%
\begin{equation}
\int_{\widehat{\Gamma }}\exp \{F(\widehat{\gamma }_{k})\}\ \Pi _{k}\left(
\mathrm{d}\widehat{\gamma }\left\vert \widehat{\eta }\right. \right) \leq
\exp \left\{ \Xi _{\delta }+\delta \sum_{j\in \partial k}F(\widehat{\eta }%
_{j})\right\} .  \label{exp1}
\end{equation}%

\end{lemma}

\textbf{Proof.} Without loss of generality we may assume that
\begin{equation*}
\delta \leq \min \left\{ A_{\Phi }-\kappa ;\text{\thinspace }A_{\chi
}-\vartheta \right\} .
\end{equation*}%
Taking into account that $Z_{\mathrm{Q}_{k}}(\widehat{\eta })\geq 1,$ cf. (%
\ref{part-bound}), and using estimate (\ref{bound1}), we obtain
\begin{multline}
\int_{\widehat{\Gamma }}\exp \{F(\widehat{\gamma }_{k})\}\ \Pi _{k}\left(
\mathrm{d}\widehat{\gamma }\left\vert \widehat{\eta }\right. \right)
\label{exp-k} \\
\leq \exp \left\{ C_{\delta }+\delta \sum_{j\in \partial k}F(\widehat{\eta }%
_{j})\right\} \int_{\widehat{\Gamma }_{k}}\exp \left\{ A_{\chi }\sum_{x\in
\gamma _{k}}\left\vert \sigma _{x}\right\vert ^{q}\right\} \ \widehat{%
\lambda }(\mathrm{d}\widehat{\gamma }_{k}).
\end{multline}%
The integral in the RHS of (\ref{exp-k}) is calculated explicitly in (\ref%
{finexp}). Then we have%
\begin{equation*}
\int_{\Gamma _{k}}\int_{S^{\gamma _{k}}}\exp \left\{ A_{\chi
}\sum_{x\in \gamma _{k}}\left\vert \sigma _{x}\right\vert ^{q}\right\}
\bigotimes_{x\in \gamma _{k}}\chi (\mathrm{d}\sigma _{x})\ \lambda (\mathrm{d%
}\gamma _{k})=\exp \left\{ z\mathcal{E}_{\chi }\right\} .
\end{equation*}%
where $\mathcal{E}_{\chi }:=\int_{S}\exp \left\{ A_{\chi }\left\vert
s\right\vert ^{q}\right\} \ \chi (\mathrm{d}s)$ is finite because of (A5).
Therefore (\ref{exp1}) holds with
\begin{equation}
\Xi _{\delta }:=C_{\delta }+z\mathcal{E}_{\chi },  \label{C-nolt}
\end{equation}%
which depends on $\mathcal{J}$ through $C_{\delta }$ and hence is
non-decreasing in $\mathcal{J}$ and $z$.
\hfill%
$\square  $

A subsequent application of Jensen's inequality to both sides in (\ref{exp1}%
) immediately implies the following estimate of Dobrushin's type (cf. \cite%
{Do70}). It states a kind of \emph{weak dependence} on boundary conditions,
which could be achieved by choosing $\delta <\mathcal{N}_{0}^{-1}.$

\begin{corollary}
\label{cor2} Under assumptions of Lemma \ref{DP1} we have the bound%
\begin{equation}
\int_{\widehat{\Gamma }}F(\widehat{\gamma }_{k})\ \Pi _{k}\left( \mathrm{d}%
\widehat{\gamma }\left\vert \widehat{\eta }\right. \right) \leq \Xi _{\delta
}+\delta \sum_{j\in \partial k}F(\widehat{\eta }_{j}).  \label{Dob}
\end{equation}
\end{corollary}

\begin{remark}
\label{rem3} By virtue of (the first inequality of) (\ref{HE-est}) and (\ref%
{finexp}) one can see that for any fixed $\kappa \in (0,A_{\Phi })$ and $%
\vartheta \in (0,A_{\chi })$
\begin{equation}
\int_{\widehat{\Gamma }}\exp \left\{ F(\widehat{\gamma }_{k})\right\} \ \Pi
_{\Lambda }\left( \mathrm{d}\widehat{\gamma }\left\vert \widehat{\eta }%
\right. \right) \leq \mathcal{C}_{k}(\Lambda ,\widehat{\eta }),\text{ \ \ }%
k\in \Lambda \in \mathcal{B}_{0}(X),  \label{expL}
\end{equation}%
where $\mathcal{C}_{k}(\Lambda ,\widehat{\eta })<\infty $ is an increasing
function of $\Lambda .$ However, this estimate is too rough for our purposes
and will be improved by more refined arguments employing the Markov property
of the specification $\Pi $, see Section \ref{sec-volume}.$ $
\end{remark}

Here and in what follows, we denote by $\mathrm{d}_{\mathrm{var}}\left( \nu
_{1},\nu _{2}\right) $ the \emph{total variation distance} between two
measures $\nu _{1}$ and $\nu _{2}$ on a $\sigma $-algebra $\mathcal{F}$,
that is,%
\begin{equation*}
\mathrm{d}_{\mathrm{var}}\left( \nu _{1},\ \nu _{2}\right) :=\mathrm{sup}%
_{A\in \mathcal{F}}\mathrm{~}\left\vert \mathfrak{\nu }_{1}(A)-\mathfrak{\nu
}_{2}(A)\right\vert .
\end{equation*}%
Our second fundamental lemma evaluates this distance between local Gibbs
states $\mu _{k}^{\widehat{\eta }}(\mathrm{d}\widehat{\gamma }_{k}):=\mu _{%
\mathrm{Q}_{k}}^{\widehat{\eta }}(\mathrm{d}\widehat{\gamma }_{k})$ and $\mu
_{k}^{\widehat{\varsigma }}(\mathrm{d}\widehat{\gamma }_{k}):=\mu _{\mathrm{Q%
}_{k}}^{\widehat{\varsigma }}(\mathrm{d}\widehat{\gamma }_{k})$ on $\mathcal{%
B}(\widehat{\Gamma }_{k})$ with boundary conditions $\widehat{\eta }$ and $%
\widehat{\varsigma }$ respectively, cf. (\ref{loc-gibbs}).

\begin{lemma}
\label{DP2} There exists a non-decreasing function $\phi (z,\mathcal{J},L)$
of $z,\mathcal{J},L>0$ such that
\begin{equation}
\mathrm{d}_{\mathrm{var}}(\mu _{k}^{\widehat{\eta }},~\mu _{k}^{\widehat{%
\varsigma }})\leq z\cdot \phi (z,\mathcal{J},L)  \label{var1}
\end{equation}%
for all $k\in \mathbb{Z}^{d}$ and any pair of boundary conditions $\widehat{%
\eta },\widehat{\varsigma }\in \Gamma (\widehat{X})$ obeying the constraint $%
\sup_{j\in \mathbb{Z}^{d}}\left\{ F(\widehat{\eta _{j}}),~F(\widehat{%
\varsigma _{j}})\right\} \leq L$.$ $
\end{lemma}

The proof is rather cumbersome and will be given in Section \ref{sec-proofs}.%
$ $

\subsection{Volume estimates \label{sec-volume}}

The aim of this section is to prove a uniform estimate on exponential
moments of the specification kernels, which in turn will be used in the
proof of Theorem \ref{existence}. For a finite subset $\mathcal{K}\subset
\mathbb{Z}^{d}$, consider the union of elementary cubes $\mathrm{Q}_{%
\mathcal{K}}:=\bigcup_{k\in \mathcal{K}}\mathrm{Q}_{k}$ (cf. (\ref{cube}))
and the corresponding cylinder set $\widehat{\mathrm{Q}}_{\mathcal{K}}=%
\mathrm{Q}_{\mathcal{K}}\times S$. Write for brevity $\Pi _{\mathcal{K}}(%
\mathrm{d}\widehat{\gamma }|\widehat{\varsigma }):=\Pi _{\mathrm{Q}_{%
\mathcal{K}}}(\mathrm{d}\widehat{\gamma }|\widehat{\varsigma })$. As usual, $%
\mathcal{K}\mathcal{\nearrow }\mathbb{Z}^{d}$ means a limit taken along any
ordered by inclusion and exhausting the whole $\mathbb{Z}^{d}$ sequence of
such sets. Our strategy will be to start from the one-point estimate (\ref%
{exp1}) and then by the consistency property (\ref{consist}) extend it to
arbitrarily large cubic domains.

\begin{theorem}
\textbf{\label{theor1}} Under assumptions of Lemma \ref{DP1} there exists a
constant $\Psi :=\Psi (\kappa ,\vartheta )<\infty $ such that the estimate
\begin{equation}
\underset{\mathcal{K}\mathcal{\nearrow }\mathbb{Z}^{d}}{\mathrm{lim~sup}}%
\int_{\widehat{\Gamma }}\exp \left\{ F(\widehat{\gamma }_{k})\right\} \ \Pi
_{\mathcal{K}}(\mathrm{d}\widehat{\gamma }|\widehat{\varsigma })\leq \Psi
\label{moment-est0}
\end{equation}%
holds for all $k\in \mathbb{Z}^{d}$ and $\widehat{\varsigma }\in \widehat{%
\Gamma }_{\mathrm{t}}(X)$.
\end{theorem}

\textbf{Proof. }Introduce the notation
\begin{equation*}
n_{k}(\mathcal{K},\widehat{\varsigma }):=\ln \int_{\widehat{\Gamma }}\exp
\left\{ F(\widehat{\gamma }_{k})\right\} ~\Pi _{\mathcal{K}}\left( \mathrm{d}%
\widehat{\gamma }|\widehat{\varsigma }\right) \geq 0,
\end{equation*}%
whereby $n_{k}(\mathcal{K},\widehat{\varsigma })=F(\widehat{\varsigma }_{k})$
if $k\notin \mathcal{K}$. An application of identity (\ref{consist}) and
inequality (\ref{exp1}) shows that for each $k\in \mathcal{K}$%
\begin{eqnarray*}
n_{k}(\mathcal{K},\widehat{\varsigma }) &=&\ln \int_{\widehat{\Gamma }}\int_{%
\widehat{\Gamma }}\exp \{F(\widehat{\gamma }_{k})\}\ \Pi _{k}\left( \mathrm{d%
}\widehat{\gamma }\left\vert \widehat{\eta }\right. \right) ~\Pi _{\mathcal{K%
}}\left( \mathrm{d}\widehat{\eta }|\widehat{\varsigma }\right)  \\
&\leq &\Xi _{\delta }+\ln \int_{\widehat{\Gamma }}\exp \left\{ \delta
\sum_{j\in \partial k}F(\widehat{\eta }_{j})\right\} ~\Pi _{\mathcal{K}%
}\left( \mathrm{d}\widehat{\eta }|\widehat{\varsigma }\right) .
\end{eqnarray*}%
Assume without loss of generality that $\delta \mathcal{N}_{0}< 1$. The
multiple H{\"{o}}lder inequality then yields%
\begin{equation*}
\int_{\widehat{\Gamma }}\prod_{j\in \partial k}\left[ \exp \{F(\widehat{\eta
}_{j})\}\right] ^{\delta }~\Pi _{\mathcal{K}}(\mathrm{d}\widehat{\eta }|%
\widehat{\varsigma })\leq \prod_{j\in \partial k}\left[ \int_{\widehat{%
\Gamma }}\exp \{F(\widehat{\eta }_{j})\}~\Pi _{\mathcal{K}}(\mathrm{d}%
\widehat{\eta }|\widehat{\varsigma })\right] ^{\delta }.
\end{equation*}%
Therefore%
\begin{equation}
n_{k}(\mathcal{K},\widehat{\varsigma })\leq \Xi _{\delta }+\delta \sum_{j\in
\mathcal{K\cap }\partial k}n_{j}(\mathcal{K},\widehat{\varsigma })+\delta
\sum_{j\in \mathcal{K}^{\mathrm{c}}\cap \partial k}F(\widehat{\varsigma }%
_{j}).  \label{ineq-n}
\end{equation}%
Fix arbitrary $k_{0}\in \mathcal{K}$ and small enough $\alpha >0$ so that $e^{\alpha
\rho }\delta \mathcal{N}_{0}<1$, where
\begin{equation*}
\rho =\sup_{k\in \mathbb{Z}^{d}}\text{\thinspace }\max_{j\in \partial k}%
\text{\textrm{\thinspace }}\left\vert j-k\right\vert \leq R+\sqrt{d}.
\end{equation*}%
Multiplying both sides of inequality (\ref{ineq-n}) by $e^{-\alpha
\left\vert k_{0}-k\right\vert }$ and taking into account that
$\left\vert k_{0}-j\right\vert -\left\vert k_{0}-k\right\vert \leq \rho $,
 we obtain the estimate%
\begin{multline}
n_{k}(\mathcal{K},\widehat{\varsigma })e^{-\alpha \left\vert
k_{0}-k\right\vert }\leq \Xi _{\delta }e^{-\alpha \left\vert
k_{0}-k\right\vert } \\
+e^{\alpha \rho }\delta \left[ \sum_{j\in \mathcal{K}\cap \partial k}n_{j}(%
\mathcal{K},\widehat{\varsigma })e^{-\alpha \left\vert k_{0}-j\right\vert
}+\sum_{j\in \mathcal{K}^{\mathrm{c}}\cap \partial k}F(\widehat{\varsigma }%
_{j})e^{-\alpha \left\vert k_{0}-j\right\vert }\right] .  \label{ineq-n1}
\end{multline}%
Thus we can see that%
\begin{multline*}
\sup_{k\in \mathcal{K}}\left\{ n_{k}(\mathcal{K},\widehat{\varsigma }%
)e^{-\alpha \left\vert k_{0}-k\right\vert }\right\}  \\
\leq \Xi _{\delta }+e^{\alpha \rho }\delta \left[ \mathcal{N}_{0}\sup_{k\in
\mathcal{K}}\left\{ n_{k}(\mathcal{K},\widehat{\varsigma })e^{-\alpha
\left\vert k_{0}-k\right\vert }\right\} +\sum_{j\in \mathcal{K}^{\mathrm{c}%
}}F(\widehat{\varsigma }_{j})e^{-\alpha \left\vert k_{0}-j\right\vert }%
\right] ,
\end{multline*}%
so that%
\begin{multline}
n_{k_{0}}(\mathcal{K},\widehat{\varsigma })\leq \sup_{k\in \mathcal{K}%
}\left\{ n_{k}(\mathcal{K},\widehat{\varsigma })e^{-\alpha \left\vert
k_{0}-k\right\vert }\right\}   \label{nk-est} \\
\leq \left( 1-e^{\alpha \rho }\delta \mathcal{N}_{0}\right) ^{-1}\left[ \Xi
_{\delta }+e^{\alpha \left( \rho +\left\vert k_{0}\right\vert \right)
}\delta \sum_{j\in \mathcal{K}^{\mathrm{c}}}F(\widehat{\varsigma }%
_{j})e^{-\alpha \left\vert j\right\vert }\right] .
\end{multline}%
It follows from (\ref{Fb}) that for any $\widehat{\varsigma }\in \widehat{%
\Gamma }_{\mathrm{t}}(X)$ we have
\begin{equation*}
\sum_{j\in \mathcal{K}^{\mathrm{c}}}F(\widehat{\varsigma }_{j})e^{-\alpha
\left\vert j\right\vert }\rightarrow 0\text{ \ as \ }\mathcal{K}\nearrow
\mathbb{Z}^{d},
\end{equation*}%
which in turn implies the bound%
\begin{equation*}
\underset{\mathcal{K}\nearrow \mathbb{Z}^{d}}{\mathrm{lim~sup~}}n_{k_{0}}(%
\mathcal{K},\widehat{\varsigma })\leq \left( 1-e^{\alpha \rho }\delta
\mathcal{N}_{0}\right) ^{-1}\Xi _{\delta }.
\end{equation*}%
Passage to the limit as $\alpha \rightarrow 0$ shows that%
\begin{equation*}
\underset{\mathcal{K}\nearrow \mathbb{Z}^{d}}{\mathrm{lim~sup~}}n_{k_{0}}(%
\mathcal{K},\widehat{\varsigma })\leq \left( 1-\delta \mathcal{N}_{0}\right)
^{-1}\Xi _{\delta }=:\Psi _{\delta },
\end{equation*}%
which completes the proof.
\hfill%
$\square $

\begin{corollary}
\label{corr1} For any domain $\Lambda \in \mathcal{B}_{0}(X)$ and $N\geq 0$,
there exists $\Psi _{\Lambda }(N)<\infty $ such that
\begin{equation*}
\limsup_{\mathcal{K}\nearrow \mathbb{Z}^{d}}\int_{\widehat{\Gamma }}F(%
\widehat{\gamma }_{\Lambda })^{N}\,\Pi _{Q_{\mathcal{K}}}(\mathrm{d}\widehat{%
\gamma }|\widehat{\varsigma })\leq \Psi _{\Lambda }(N),
\end{equation*}%
which holds uniformly for all $\widehat{\varsigma }\in \widehat{\Gamma }_{%
\mathrm{t}}(X)$.
\end{corollary}

\section{Existence of Gibbs measures \label{sec-exist}}

In this section, we use the estimates obtained in Section\ \ref{ME} in order
to prove that, for any $\widehat{\eta }\in \widehat{\Gamma }_{\mathrm{t}}(X)$%
, the family of Gibbsian specification kernels $\left\{ \Pi _{\Lambda
}\left( \cdot |\widehat{\eta }\right) ,\ \Lambda \in \mathcal{B}%
_{0}(X)\right\} $ contains a cluster point.

\begin{definition}
\label{equi} (cf. \cite[Def. 4.6]{Geor}) We say that a sequence of
probability measures $\left\{ \mu _{m}\right\} _{m\in \mathbb{N}}$ on $%
\widehat{\Gamma }(X)$ is locally equicontinuous (LEC) if for any $\Lambda
\in \mathcal{B}_{0}(X)$ and any $\left\{ B_{n}\right\} _{n\in \mathbb{N}%
}\subset \mathcal{B}_{\Lambda }(\widehat{\Gamma })$ with $B_{n}\searrow
\emptyset $ as $n\rightarrow \infty $, we have%
\begin{equation}
\underset{n\rightarrow \infty }{\mathrm{lim}}\mathrm{\ }\underset{m\in
\mathbb{N}}{\mathrm{lim~sup}}\ \mu _{m}\left( B_{n}\right) =0.
\label{equicont0}
\end{equation}
\end{definition}

We equip the space $\mathcal{P}(\widehat{\Gamma })$ of probability measures
on $\widehat{\Gamma }(X)$ with the topology of \emph{local set convergence},
which is defined as the coarsest topology making the evaluation map $\mu
\rightarrow \mu (B)$ continuous for each $B\in \mathcal{F}_{0}:=\mathcal{B}%
_{0}(\widehat{\Gamma })$. This topology (which is Hausdorff but \emph{not }%
metrizable) is well suited to the study of local interactions (i.e., those
having finite range as in assumption (A1)). In particular,
\begin{equation}
\mu _{m}\overset{\mathrm{loc}}{\rightarrow }\mu \ \ \text{iff \ }\mu
_{m}(B)\rightarrow \mu (B)\ \ \text{as }m\rightarrow \infty ,\ \ \forall
B\in \mathcal{F}_{0}.  \label{equi1}
\end{equation}%
The latter is equivalent to claiming that
\begin{equation}
\int_{\widehat{\Gamma }}f\,\mathrm{d}\mu _{m}\rightarrow \int_{\widehat{%
\Gamma }}f\,\mathrm{d}\mu \text{ \ as }m\rightarrow \infty ,  \label{equi2}
\end{equation}%
for all bounded $\mathcal{F}_{0}$-measurable functions $f:$ $\widehat{\Gamma
}(X)\rightarrow \mathbb{R}$. Observe that the local set convergence is
equivalent to convergence in the space $[0,1]^{\mathcal{F}_{0}}$.

\begin{theorem}
\label{equicont1}(cf. \cite[Prop. 4.9]{Geor}) Any LEC sequence $\left\{ \mu
_{m}\right\} _{m\in \mathbb{N}}\subset \mathcal{P}(\widehat{\Gamma })$ has
at least one cluster point, which is a probability measure on $\widehat{%
\Gamma }(X)$.
\end{theorem}

\textbf{Sketch of the proof.} It is straightforward that the family $\left\{
\mu _{m}\right\} _{m\in \mathbb{N}}$ contains a cluster point $\mu $ as an
element of the compact space $[0,1]^{\mathcal{F}_{0}}$, and $\mu $ is an
additive function on $\mathcal{F}_{0}$. The LEC property (\ref{equicont0})
implies that $\mu _{\Lambda }:=p_{\Lambda }^{\ast }\mu $ is $\sigma $%
-additive on each $\mathcal{B}(\widehat{\Gamma }_{\Lambda })$. Thus $\left\{
\mu _{\Lambda }\right\} _{\Lambda \in \mathcal{B}_{0}(X)}$ forms a
consistent (w.r.t. projective maps (\ref{pl2})) family of measures and by
the corresponding version of the Kolmogorov theorem (see \cite[Theorem V.3.2
]{Par}) generates a probability measure on $\mathcal{B}(\widehat{\Gamma })$
(which obviously coincides with $\mu $).
\hfill%
$\square  $

\begin{remark}
\label{rem_5} It follows from \cite[Prop. 4.15]{Geor} that, although the
topology of $\mathcal{P}(\widehat{\Gamma })$ is not metrizable, for each
(topological)
cluster point $\mu $ there exists a subsequence $\{\mu _{m_{j}}\}_{j\in
\mathbb{N}}$ such that $\mu _{m_{j}}\overset{\mathrm{loc}}{\rightarrow }\mu $
as $j\rightarrow \infty $.
\end{remark}

Let now $\left\{ \mathcal{K}_{m}\right\} _{m\in \mathbb{N}}$ be any
increasing sequence of finite subsets of $\mathbb{Z}^{d}$ such that $%
\mathcal{K}_{m}\nearrow \mathbb{Z}^{d}\mathcal{\ }$and hence $\mathrm{Q}_{%
\mathcal{K}_{m}}:=\bigcup_{j\in \mathcal{K}_{m}}\mathrm{Q}_{j}\nearrow X$ as
$m\rightarrow \infty $, and introduce notation $\Lambda _{m}:=\Lambda _{%
\mathcal{K}_{m}}$ and $\Pi _{m}:=\Pi _{\Lambda _{\mathcal{K}_{m}}}$.

\begin{theorem}
\label{equicont2}For any $\widehat{\varsigma }\in \widehat{\Gamma }_{\mathrm{%
t}}(X)$ the family $\left\{ \Pi _{m}\left( \mathrm{d}\widehat{\gamma }|%
\widehat{\varsigma }\right) \right\} _{m\in \mathbb{N}}$ is LEC.
\end{theorem}

\textbf{Proof. }Fix $\Lambda \in \mathcal{B}_{0}(X)$ and $\left\{
B_{n}\right\} _{n\in \mathbb{N}}\subset \mathcal{B}_{\Lambda }(\widehat{%
\Gamma })$ as in Definition \ref{equi}. It is sufficient to prove that $%
\forall \varepsilon >0$ there exist integers $m_{0}$ and $n_{0}$ such that $%
\Pi _{m}(B_{n}|\widehat{\varsigma })\leq \varepsilon $ for any $m\geq m_{0}$
and $n\geq n_{0}$.$ $

To this end, for $T>0$ let us consider the set
\begin{equation*}
\widehat{\Gamma }_{T}:=\left\{ \widehat{\gamma }\in \widehat{\Gamma }(X):%
\text{ }F\left( \widehat{\gamma }_{\Lambda _{R}}\right) =\kappa N(\gamma
_{\Lambda _{R}})^{P}+\vartheta \sum_{x\in \gamma _{\Lambda _{R}}}\left\vert
\sigma _{x}\right\vert ^{q}\leq T\right\}
\end{equation*}%
(where $\Lambda _{R}$ was defined in Section \ref{notations}) and estimate
the corresponding measures of $B_{n}\cap \widehat{\Gamma }_{T}$ and $%
B_{n}\cap \lbrack \widehat{\Gamma }_{T}]^{\mathrm{c}}$ separately. Observe
(by analogy with (\ref{DP_bounds}) and (\ref{DP3})) that for any $1\leq
p\leq P$ and $1\leq r\leq q$
\begin{equation*}
\sup_{\widehat{\gamma }\in \widehat{\Gamma }_{T}}\left\{ N(\gamma _{\Lambda
_{R}})^{p};\text{ }\sum_{x\in \gamma _{\Lambda _{R}}}\left\vert \sigma
_{x}\right\vert ^{r}\right\} \leq \frac{T}{\max \left\{ \kappa ;\vartheta
\right\} }.
\end{equation*}%
Using bound (\ref{HE-est}) we then see that there exists a constant $%
c_{\Lambda }(T)$ such that
\begin{equation}
\mathbf{1}_{\widehat{\Gamma }_{T}}\left( \widehat{\eta }_{\Lambda }\cup
\widehat{\gamma }_{\Lambda ^{c}}\right) \mathrm{\exp }\left\{ -H_{\Lambda }(%
\widehat{\eta }_{\Lambda }|\widehat{\gamma })\right\} \leq c_{\Lambda }(T).
\label{est1}
\end{equation}%
uniformly for all $\widehat{\gamma },\widehat{\eta }\in \widehat{\Gamma }%
(X) $.

Next, write%
\begin{equation*}
\Pi _{m}\left( B_{n}\left\vert \widehat{\varsigma }\right. \right) =\Pi
_{m}(B_{n}\cap \lbrack \widehat{\Gamma }_{T}]^{\mathrm{c}}|\widehat{%
\varsigma })+\Pi _{m}(B_{n}\cap \widehat{\Gamma }_{T}\left\vert \widehat{%
\varsigma }\right. ).
\end{equation*}%
According to Chebyshev's inequality applied to the measure $\Pi _{m}\left(
\mathrm{d}\widehat{\gamma }|\widehat{\varsigma }\right) $ on $\widehat{%
\Gamma }(X)$ we have%
\begin{equation*}
\Pi _{m}\left( \left\{ \widehat{\gamma }:f\left( \widehat{\gamma }\right)
\geq T\right\} |\widehat{\varsigma }\right) \leq T^{-2}\int_{\widehat{\Gamma
}}\left\vert f(\widehat{\gamma })\right\vert ^{2}~\Pi _{m}\left( \mathrm{d}%
\widehat{\gamma }|\widehat{\varsigma }\right)
\end{equation*}%
for any $T>0$ and $f\in L^{2}(\widehat{\Gamma },\ \Pi _{m}\left( \mathrm{d}%
\widehat{\gamma }|\widehat{\varsigma }\right) )$. Setting $f\left( \widehat{%
\gamma }\right) =F\left( \widehat{\gamma }_{\Lambda _{R}}\right) $ we
obtain, cf. Corollary \ref{corr1},%
\begin{equation}
\Pi _{m}(B_{n}\cap \lbrack \widehat{\Gamma }_{T}]^{\mathrm{c}}\left\vert
\widehat{\varsigma }\right. )\leq \Pi _{m}\left( [\widehat{\Gamma }_{T}]^{%
\mathrm{c}}\left\vert \widehat{\varsigma }\right. \right) \leq \varepsilon /2
\label{cheb}
\end{equation}%
for any $\varepsilon >0$ and $T$ greater than some $T(\varepsilon )$.

On the other hand, there exists $m_{0}$ such that $\Lambda _{m}\supset
\Lambda $ for $m\geq m_{0}$. For all such $m$, it follows from (\ref{specif0}%
) and the consistency property (\ref{consist}) of the specification $\Pi $
that
\begin{equation}
\Pi _{m}\left( B_{n}\cap \widehat{\Gamma }_{T}\left\vert \widehat{\varsigma }%
\right. \right) =\int_{\widehat{\Gamma }}\left[ \int_{\widehat{\Gamma }}%
\mathbf{1}_{B_{n}\cap \widehat{\Gamma }_{T}}\left( \widehat{\eta }_{\Lambda
}\cup \widehat{\gamma }_{\Lambda ^{c}}\right) \text{~}\Pi _{\Lambda }\left(
\mathrm{d}\widehat{\eta }|\widehat{\gamma }\right) \right] ~\Pi _{m}\left(
\mathrm{d}\widehat{\gamma }|\widehat{\varsigma }\right) .  \label{Pim}
\end{equation}%
Since $B_{n}\downarrow \emptyset $ as $n\rightarrow \infty $, by (\ref%
{part-bound}) and (\ref{est1}) we obtain%
\begin{equation*}
\int_{\widehat{\Gamma }}\mathbf{1}_{B_{n}\cap \widehat{\Gamma }_{T}}\left(
\widehat{\eta }_{\Lambda }\cup \widehat{\gamma }_{\Lambda ^{c}}\right) ~\Pi
_{\Lambda }\left( \mathrm{d}\widehat{\eta }\left\vert \widehat{\gamma }%
\right. \right) \leq c_{\Lambda }(T)\widehat{\lambda }_z(B_{n})<\varepsilon /2
\end{equation*}%
for $n$ greater than some $n(\varepsilon ,T)$. Hence, the right-hand side in
(\ref{Pim}) does not exceed $\varepsilon /2$ as well. Combining this with
estimate (\ref{cheb}) we can see that $\forall \varepsilon >0$ and $m\geq
m_{0}$, $n\geq n_{0}=n(\varepsilon ,T(\varepsilon ))$ it holds%
\begin{equation*}
\Pi _{m}\left( B_{n}|\widehat{\varsigma }\right) \leq \varepsilon
/2+\varepsilon /2=\varepsilon ,
\end{equation*}%
which completes the proof.
\hfill%
$\square  $

Now we are in a position to prove our first main result.$ $

\textbf{Proof of Theorem \ref{existence}.} \emph{(i) Existence:} It follows
from Theorems \ref{equicont1} and \ref{equicont2} that for any $\widehat{%
\varsigma }\in \Gamma ^{\mathrm{t}}$ the family $\left\{ \Pi _{m}\left(
\mathrm{d}\widehat{\gamma }\left\vert \widehat{\varsigma }\right. \right)
\right\} _{m\in \mathbb{N}}$ has a cluster point $\mu =\mu (\widehat{%
\varsigma })\in \mathcal{P}(\widehat{\Gamma })$. Therefore by Remark \ref%
{rem_5} there exists a subsequence $\Lambda _{m_{j}},\ j\in \mathbb{N}$,
such that%
\begin{equation}
\lim_{j\rightarrow \infty }~\Pi _{m_{j}}\left( B\left\vert \widehat{%
\varsigma }\right. \right) =\mu (B),\ \ B\in \mathcal{B}_{0}(\widehat{\Gamma
}).  \label{setwise_conv}
\end{equation}%
Let us check that $\mu $ solves the DLR\ equation (\ref{DLR}) for all $%
\Lambda \in \mathcal{B}_{0}(X)$ and $B\in \mathcal{B}_{0}(\widehat{\Gamma })$%
. As the interaction has finite range, the function $\widehat{\gamma }%
\mapsto \Pi _{\Lambda }\left( B\left\vert \widehat{\gamma }\right. \right) $
is $\mathcal{B}_{0}(\widehat{\Gamma })$-measurable. Using (\ref{equi2}) and
the consistency property (\ref{consist}) of the specification $\Pi $, we
thus can pass to the limit
\begin{multline*}
\int_{\widehat{\Gamma }}\Pi _{\Lambda }\left( B\left\vert \widehat{\gamma }%
\right. \right) \,\mu \left( \mathrm{d}\widehat{\gamma }\right)
=\lim_{j\rightarrow \infty }\int_{\widehat{\Gamma }}\Pi _{\Lambda }\left(
B\left\vert \widehat{\gamma }\right. \right) \,\Pi _{m_{j}}\left( \mathrm{d}%
\widehat{\gamma }\left\vert \widehat{\varsigma }\right. \right) \\
=\lim_{j\rightarrow \infty }~\Pi _{m_{j}}\left( B\left\vert \widehat{%
\varsigma }\right. \right) =\mu (B)
\end{multline*}%
and conclude that $\mu \in \mathcal{G}.$ Finally, by (\ref{moment-est0}) and
Beppo Levi's monotone convergence theorem we see that
\begin{multline*}
\int_{\widehat{\Gamma }}\sum_{k\in \mathbb{Z}^{d}}e^{-\alpha \left\vert
k\right\vert }F(\widehat{\gamma }_{k})\,\mu \left( \mathrm{d}\widehat{\gamma
}\right) =\lim_{K,L\rightarrow \infty }\lim_{j\rightarrow \infty
}\sum_{|k|\leq K}e^{-\alpha \left\vert k\right\vert }\int_{\widehat{\Gamma }%
}\{F(\widehat{\gamma }_{k})\wedge L\}\,\Pi _{m_{j}}\left( \mathrm{d}\widehat{%
\gamma }\left\vert \widehat{\varsigma }\right. \right) \\
\leq \sum_{k\in \mathbb{Z}^{d}}e^{-\alpha \left\vert k\right\vert
}\limsup_{j\rightarrow \infty }\int_{\widehat{\Gamma }}F(\widehat{\gamma }%
_{k})\,\Pi _{m_{j}}\left( \mathrm{d}\widehat{\gamma }\left\vert \widehat{%
\varsigma }\right. \right) \leq \Psi \sum_{k\in \mathbb{Z}^{d}}e^{-\alpha
\left\vert k\right\vert }<\infty
\end{multline*}%
for all $\alpha >0$, which by (\ref{Fb}) implies that $\mu (\widehat{\Gamma }%
_{\mathrm{t}}(X))=1$ $ $so that $\mu \in \mathcal{G}^{\mathrm{t}}$.

\emph{(ii) A priori estimate (\ref{moment-est}). }Consider an arbitrary $\mu
\in \mathcal{G}^{\mathrm{t}}$ (not necessarily given by the limit transition
above). With the help of (\ref{DLR}), Theorem \ref{theor1} and Fatou's lemma
we have
\begin{multline*}
\int_{\widehat{\Gamma }_{\mathrm{t}}}\exp \{F(\widehat{\gamma }_{k})\wedge
L\}\,\mu \left( \mathrm{d}\widehat{\gamma }\right) =\lim_{\mathcal{K}%
\nearrow \mathbb{Z}^{d}}\int_{\widehat{\Gamma }_{\mathrm{t}}}\int_{\widehat{%
\Gamma }}\exp \{F(\widehat{\gamma }_{k})\wedge L\}\,\Pi _{\mathcal{K}%
}\left( \mathrm{d}\widehat{\gamma }\left\vert \widehat{\varsigma }\right.
\right) \mu \left( \mathrm{d}\widehat{\varsigma }\right) \\
\leq \int_{\widehat{\Gamma }_{\mathrm{t}}}\left[ \limsup_{\mathcal{K}%
\nearrow \mathbb{Z}^{d}}\int_{\widehat{\Gamma }}\exp \{F(\widehat{\gamma }%
_{k})\wedge L\}\,\Pi _{\mathcal{K}}\left( \mathrm{d}\widehat{\gamma }%
\left\vert \widehat{\varsigma }\right. \right) \right] \mu \left( \mathrm{d}%
\widehat{\varsigma }\right) \leq \Psi
\end{multline*}%
for any $k\in \mathbb{Z}^{d}$ and $L>0$, where $\Psi >0$ is the same
constant as in (\ref{moment-est0}). By Levi's theorem this implies the bound%
\begin{equation*}
\int_{\widehat{\Gamma }}\exp \{F(\widehat{\gamma }_{k})\}\,\mu \left(
\mathrm{d}\widehat{\gamma }\right) =\lim_{L\rightarrow \infty }\int_{%
\widehat{\Gamma }_{\mathrm{t}}}\exp \{F(\widehat{\gamma }_{k})\wedge
L\}\,\mu \left( \mathrm{d}\widehat{\gamma }\right) \leq \Psi ,
\end{equation*}%
and (\ref{moment-est}) is proved.
\hfill%
$\square  $

\begin{remark}
\label{cor4} A standard application of the Borel--Cantelli lemma to the
moment bound (\ref{moment-est}) yields the following improved support
property for any $\mu \in \mathcal{G}^{\mathrm{t}}$. Indeed, under the
conditions of Theorem \ref{existence} all $\mu \in \mathcal{G}^{\mathrm{t}}$
are carried by the set
\begin{equation}
\widehat{\Gamma }_{\mathrm{s}}(X)=\left\{ \widehat{\gamma }\in \widehat{%
\Gamma }(X):\ \sup_{k\in \mathbb{Z}^{d}}\ \left[ N(\gamma
_{k})^{P}+ \sum_{x\in \gamma _{k}}\left\vert \sigma _{x}\right\vert
^{q}\right] \cdot \left[ \log \left( 1+|k|\right) \right] ^{-1}<\infty
\right\} ,  \label{gs}
\end{equation}%
which is smaller than $\widehat{\Gamma }_{\mathrm{t}}(X)$, cf. (\ref{temper}%
) and (\ref{Fb}).
\end{remark}

\begin{remark}
\label{rem_4} Let us consider a special case when all the potentials are
non-negative, i.e., $\Phi \left( x,y\right) \geq 0$ and $W_{xy}(s,t)\geq 0.$
This would make superfluous the superstability assumptions (A3) and (A6).
Indeed, in this case we can use the control functional%
\begin{equation*}
\tilde{F}(\widehat{\gamma }):=\kappa N(\gamma )+\vartheta \sum_{x\in \gamma
}\left\vert \sigma _{x}\right\vert ^{q},\ \ \widehat{\gamma }=(\gamma
,\sigma ),
\end{equation*}%
instead of (\ref{Fk}), with arbitary fixed $\kappa >0$ and $\vartheta \in
(0,A_{\chi })$. Then we have the estimate
\begin{multline}
\int_{\widehat{\Gamma }_{k}}\exp \{\tilde{F}(\widehat{\gamma }_{k})\}\ \mu
_{k}^{\widehat{\eta }}\left( \mathrm{d}\widehat{\gamma }_{k}\right) \leq
\int_{\widehat{\Gamma }_{k}}\exp \{\tilde{F}(\widehat{\gamma }_{k})\}\
\widehat{\lambda }_z\left( \mathrm{d}\widehat{\gamma }_{k}\right)  \label{Fn}
\\
=\exp \left\{ ze^{\kappa }\int_{S}e^{\vartheta \left\vert s\right\vert
^{q}}\ \chi (\mathrm{d}s)\right\} <\infty ,
\end{multline}%
which holds uniformly for all $\widehat{\eta }\in \widehat{\Gamma }(X)$ and $%
k\in \mathbb{Z}^{d}$, cf. (\ref{finexp}). This enables us to mimic the proof
of Theorem \ref{existence}\textbf{\ }and construct in this way a Gibbs
measure $\mu \in \mathcal{G}$ obeying the \emph{a priori} bound $%
\sup_{k}\int_{\widehat{\Gamma }}\exp \{\tilde{F}(\widehat{\gamma }_{k})\}\
\mu \left( \mathrm{d}\widehat{\gamma }\right) <\infty .$
\end{remark}

\section{Uniqueness of Gibbs measures\label{sec-uniq}}

The aim of this section is to prove Theorem \ref{uniqueness}. First we will
develop the lattice representation of our model, in order to use the
abstract Dobrushin--Pechersky uniqueness criterion.

\subsection{Lattice representation of the model\label{sec-lattice}}

Let $\mathcal{Q}:=\widehat{\Gamma }_{\mathrm{Q}_{0}}$, where $\mathrm{Q}_{0}$
is the elementary cube centered at the origin, cf. (\ref{cube}). Recall that
$(\mathcal{Q},\mathcal{B}(\mathcal{Q}))$ is a standard Borel space and fix
the Lebesgue-Poisson measure $\widehat{\lambda }_z$ thereon. Consider the
product space $\mathcal{A}:=\mathcal{Q}^{\mathbb{Z}^{d}}=\prod_{k\in \mathbb{%
Z}^{d}}\mathcal{Q}_{k}$, $\mathcal{Q}_{k}:=\mathcal{Q}$, and endow it with
the product topology and the corresponding Borel $\sigma $-algebra $\mathcal{%
B}(\mathcal{A})$. Elements of $\mathcal{A}$, to be called \emph{lattice
configurations}, are infinite sequences $\overline{\alpha }:=(\alpha
_{k})_{k\in \mathbb{Z}^{d}}$ with $\alpha _{k}\in \mathcal{Q}$. By
construction, $\mathcal{B}(\mathcal{A})$ is generated by cylinder sets
\begin{equation}
A_{b_{1},...,b_{m}}^{k_{1},...,k_{m}}:=\left\{ \overline{\alpha }\in
\mathcal{A}:~\alpha _{k_{1}}\in b_{1},\ldots ,\alpha _{k_{m}}\in
b_{m}\right\}  \label{Ak}
\end{equation}%
with all possible choices of $k_{i}\in \mathbb{Z}^{d},$ $b_{i}\in \mathcal{B}%
(\mathcal{Q})$ and $1\leq i\leq m\in \mathbb{N}$.

\begin{remark}
Observe that in our notations $\mathcal{Q}_{k}$ is the $k$-th copy of $%
\mathcal{Q}=\widehat{\Gamma }_{\mathrm{Q}_{0}}$, so that $\mathcal{Q}%
_{k}\neq \widehat{\Gamma }_{\mathrm{Q}_{k}}$. These spaces are isomorphic
via the translation by $k$.
\end{remark}

Define the map%
\begin{equation}
\mathbb{T}:\widehat{\Gamma }(X)\ni \widehat{\gamma }\longmapsto \mathbb{T}(%
\widehat{\gamma })=\overline{\alpha }\in \mathcal{A}  \label{T1}
\end{equation}%
where $\overline{\alpha }:=(\alpha _{k})_{k\in \mathbb{Z}^{d}}$ with $\alpha
_{k}=\widehat{\gamma }_{k}-k\in \widehat{\Gamma }_{\mathrm{Q}_{0}}$. Here we
write $$\widehat{\eta }-a:=\left\{ ~...,~(x-a,s),~...~\right\} $$ for a marked
configuration $\widehat{\eta }=\left\{ ~...,~(x,s),~...~\right\} \in
\widehat{\Gamma }(X)$ and $a\in X$. Moreover, for any $B\in \widehat{\Gamma }%
(X)$ we define the shifted set $B-a$ constituted by all configurations $%
\widehat{\eta }-a$ with $\widehat{\eta }\in B$.

\begin{lemma}
\label{meas} $\mathbb{T}\mathbf{:~}\widehat{\Gamma }(X)\rightarrow \mathcal{A%
}$ is a measurable bijection.
\end{lemma}

\noindent \textbf{Proof.} The map $\mathbb{T}$ is clearly one-to-one by its
construction. The inverse map $\mathbb{T}^{-1}$ acts as%
\begin{equation}
\mathbb{T}^{-1}:\mathcal{A}\ni \overline{\alpha }\longmapsto \mathbb{T}^{-1}(%
\overline{\alpha })=\widehat{\gamma }\in \widehat{\Gamma }(X)  \label{T2}
\end{equation}%
where $\widehat{\gamma }:=\bigcup_{k\in \mathbb{Z}^{d}}(\alpha _{k}+k)$. To
establish the measurability of $\mathbb{T}$ it is sufficient to consider
cylinder sets of the form (\ref{Ak}). Then
\begin{equation}
\mathbb{T}^{-1}\left( A_{b_{1},...,b_{m}}^{k_{1},...,k_{m}}\right)
=\bigcap_{1\leq i\leq m}B_{(k_{i},b_{i})}\in \mathcal{B}(\widehat{\Gamma }),
\label{T3}
\end{equation}%
where $B_{(k,b)}\in \mathcal{B}_{0}(\widehat{\Gamma })$ is defined for each $%
k\in \mathbb{Z}^{d}$ and $b\in \mathcal{B}(\mathcal{Q})$ as follows:%
\begin{equation*}
B_{(k,b)}:=\left\{ \widehat{\gamma }\in \widehat{\Gamma }(X):\text{ }%
\widehat{\gamma }_{k}\in \tilde{b}_{k}\right\} ,\text{ \ }\tilde{b}%
_{k}:=b+k\in \mathcal{B}(\widehat{\Gamma }_{\mathrm{Q}_{k}}).
\end{equation*}%
Furthermore, observe that such sets on the right-hand side in (\ref{T3})
generate the whole $\mathcal{B}(\widehat{\Gamma })$, which means the
measurability of $\mathbb{T}^{-1}$ as well.
\hfill%
$\square  $

Thus, for any $\mu \in \mathcal{P}(\widehat{\Gamma })$ we can define its
push-forward image $\mathbb{T}_{\ast }\mu \in \mathcal{P}(\mathcal{A})$,
where $\mathcal{P}(\mathcal{A})$ is the set of all probability measures on $%
\mathcal{A}$.

\begin{lemma}
\label{lattice}The map $\mathbb{T}_{\ast }:\mathcal{P}(\widehat{\Gamma }%
)\rightarrow \mathcal{P}(\mathcal{A})$ is injective.
\end{lemma}

\noindent \textbf{Proof.} Let $\mu ,\nu \in \mathcal{P}(\widehat{\Gamma })$
and $\mu \neq \nu $. Then there exists $B\in \mathcal{B}(\widehat{\Gamma })$
such that $\mu (B)\neq \nu (B)$. By Lemma \ref{meas}, $A:=\mathbb{T}(B)\in
\mathcal{B}(\mathcal{A})$ and $\mathbb{T}^{-1}(A)=B$. Thus $\mathbb{T}_{\ast
}\mu (A)=\mu (\mathbb{T}^{-1}(A))\neq \nu (\mathbb{T}^{-1}(A))=\mathbb{T}%
_{\ast }\nu (A)$, and the statement is proved.
\hfill%
$\square  $

Define a family of one-point states $\mathfrak{M}=\left\{ \mathfrak{m}_{k}^{%
\overline{\alpha }}:\ k\in \mathbb{Z}^{d},\text{ }\overline{\alpha }\in
\mathcal{A}\right\} $ by the formula%
\begin{equation}
\mathfrak{m}_{k}^{\overline{\alpha }}(b):=\mu _{k}^{\mathbb{T}^{-1}%
\overline{\alpha }}\left( b +k\right) ,\ \ b\in \mathcal{B}\left(
\mathcal{Q}\right) ,  \label{mb}
\end{equation}%
where $\mu _{k}:=\mu _{\mathrm{Q}_{k}}$ is the local Gibbs state of the
initial model given by (\ref{loc-gibbs}). The corresponding one-point
specification $\mathfrak{P}=\left\{ \mathfrak{p}_{k}^{\overline{\alpha }}:\
k\in \mathbb{Z}^{d},\text{ }\overline{\alpha }\in \mathcal{A}\right\} $ is
constituted by probability kernels%
\begin{equation*}
\mathcal{A}\times \mathcal{B}(\mathcal{A})\ni (\overline{\alpha },A)\mapsto
\mathfrak{p}_{k}^{\overline{\alpha }}(A):=\Pi _{k}\left( \mathbb{T}^{-1}A\left\vert
\mathbb{T}^{-1}\overline{\alpha }\right. \right) ,
\end{equation*}
cf. (\ref{specif}). It is clear that $\mathfrak{m}_{k}^{\overline{\alpha }%
}\in \mathcal{P}(\mathcal{Q}_{k})$ coincides with the projection of $%
\mathfrak{p}_{k}^{\overline{\alpha }}\in \mathcal{P}(\mathcal{A})$ onto the $%
k$-th component of the product space $\mathcal{A}$.

\begin{lemma}
\label{L-m}For any $k\in \mathbb{Z}^{d}$ and $\overline{\alpha },\ \overline{%
\alpha }^{\prime }\in \mathcal{A}$ we have the following statements:

\begin{enumerate}
\item[(i)] Measure $\mathfrak{m}_{k}^{\overline{\alpha }}$ has the form%
\begin{equation*}
\mathfrak{m}_{k}^{\overline{\alpha }}(\mathrm{d}\beta
)={\cal Z}^{-1}e^{-{\cal H}_{k}(\beta \left\vert \overline{\alpha }\right. )}~\widehat{%
\lambda }(\mathrm{d}\beta ),
\end{equation*}%
where ${\cal H}_{k}(\beta \left\vert \overline{\alpha }\right. ):=H_{\mathrm{Q}%
_{k}}(\beta +k\left\vert \mathbb{T}^{-1}\overline{\alpha }\right. )$, $\beta
\in \mathcal{Q}:=\widehat{\Gamma }_{\mathrm{Q}_{0}}$ and ${\cal Z}:=Z_{\mathrm{Q}%
_{k}}(\mathbb{T}^{-1}\overline{\alpha })$ is the normalizing factor (cf. (%
\ref{norm_fact})).

\item[(ii)] Assume that $\overline{\alpha }_{\partial k}=\overline{\alpha }%
_{\partial k}^{\prime }$, where $\partial k$ is defined in Sec. \ref%
{notations}. Then $\mathfrak{m}_{k}^{\overline{\alpha }}=\mathfrak{m}_{k}^{%
\overline{\alpha }^{\prime }}$ (Markovian property).
\end{enumerate}
\end{lemma}

\noindent \textbf{Proof.} The statement immediately follows from the
definition of measure $\mathfrak{m}_{k}^{\overline{\alpha }}$ and energy
function $H_{k}$, cf. (\ref{HL1}), and the translation invariance of the
Lebesgue-Poisson measure $\widehat{\lambda }_z$.$ $
\hfill%
$\square $

We denote by $\mathcal{M}(\mathfrak{P})$ the set of probability measures $%
\varpi \in \mathcal{P(A)}$ which are consistent with the singleton
specification $\mathfrak{P}$, that is,
\begin{equation}
\int_{\mathcal{A}}\mathfrak{p}_{k}^{\bar{\alpha}}(A)~\varpi (\mathrm{d}%
\overline{\alpha })=\varpi (A),\ \text{\ }k\in \mathbb{Z}^{d}\text{, }A\in
\mathcal{B}(\mathcal{A}).  \label{cons}
\end{equation}%
For a measurable non-negative function $h:\mathcal{Q}\rightarrow \mathbb{R}$
define the subset $\mathcal{M}_{h}(\mathfrak{P})$ of those $\varpi \in
\mathcal{M}(\mathfrak{P})$ that satisfy the bound%
\begin{equation}
\sup_{k\in \mathbb{Z}^{d}}\int_{\mathcal{A}}h(\alpha _{k})~\varpi (\mathrm{d}%
\overline{\alpha })<\infty .  \label{moments}
\end{equation}

\begin{lemma}
\label{lattice1} Let $\mu \in \mathcal{G}^{\mathrm{t}}$. Then $\mathbb{T}%
_{\ast }\mu \in \mathcal{M}_{h_{F}}(\mathfrak{P})$ with $h_{F}=F_{\lceil
\mathcal{Q}}$, where $F$ is defined by formula (\ref{Fk}).
\end{lemma}

\noindent \textbf{Proof.} The consistency property (\ref{cons}) and bound (%
\ref{moments}) follow directly from the DLR equation (\ref{DLR}) and
estimate (\ref{moment-est}), respectively.$ $
\hfill%
$\square $

The next statement is crucial for our approach.

\begin{proposition}
\label{lattice2}We have $N\left( \mathcal{G}^{\mathrm{t}}\right) \leq
N\left( \mathcal{M}_{h_{F}}(\mathfrak{P})\right) $.
\end{proposition}

\noindent \textbf{Proof.} Follows directly from Lemmas \ref{lattice} and \ref%
{lattice1}. \smallskip
\hfill%
$\square $

Thus, in order to show that $\mathcal{G}^{\mathrm{t}}$ contains at most one
element, it is sufficient to prove that $\mathcal{M}_{h_{F}}(\mathfrak{P})$
does so.$ $

The uniqueness in question will be studied with the help of the
Dobrushin--Pechesky criterion for lattice Gibbs states, extending
Dobrushin's famous criterion \cite{Do70} to the case of non-compact spins.
This abstract result originally appeared in \cite{DoP}, see also \cite[%
Theorem 2.6]{CKKP} for its further developments and \cite[Theorem 3]{BP},
\cite[Theorem 4]{PZ} resp. \cite{Pasurek} for applications to some models of
interacting particle systems (both in the continuum and on a lattice). More
precisely, we will use the following adaptation of the Dobrushin--Pechesky
criterion to our setting. $ $

\begin{theorem}[Uniqueness Criterion]
\label{Thm_DPuniqueness} There exist a positive threshold value $\delta
_{\ast }:=\delta _{\ast }(d,R)<1$ and a function $L^{\ast }:\mathbb{R}%
_{+}^{3}\rightarrow \left( 0,\infty \right) $ such that $N\left( \mathcal{M}%
_{h}(\mathfrak{P})\right) \leq 1$ provided the family $\mathfrak{M}$ of
one-point local Gibbs states satisfies the following two conditions:

\begin{itemize}
\item[\textbf{(DP-1)}] There exist constants $\delta <\delta _{\ast }$ and $\Xi >0$ such that%
\begin{equation*}
\int_{\mathcal{Q}}h(\beta )~\mathfrak{m}_{k}^{\bar{\alpha}}(\mathrm{d}\beta
)\leq \Xi +\delta \sum_{j\in \partial k}h(\alpha _{j})
\end{equation*}%
for any $k\in \mathbb{Z}^{d}$ and all boundary conditions $\bar{\alpha}\in
\mathcal{A}$.

\item[\textbf{(DP-2)}] There exists a constant $\ell <\mathcal{N}_{0}^{-1}$
such that%
\begin{equation*}
\mathrm{d}_{\mathrm{var}}\left( \mathfrak{m}_{k}^{\overline{\alpha }},~%
\mathfrak{m}_{k}^{\overline{\alpha }^{\prime }}\right) <\ell
\end{equation*}%
for any $k\in \mathbb{Z}^{d}$ and all boundary conditions $\overline{\alpha }%
,\overline{\alpha }^{\prime }\in \mathcal{A}$ obeying the constraint%
\begin{equation}
\sup_{j\in \mathbb{Z}^{d}}~\{h(\overline{\alpha }_{j});~h(\overline{\alpha }%
_{j}^{\prime })\}\leq L^{\ast }(\Xi ,\delta ,\ell ).  \label{CC-bc}
\end{equation}
\end{itemize}
\end{theorem}

\begin{remark}
The original result is more refined in that precise threshold values $\delta
_{\ast }$ and $L^{\ast }(\Xi ,\delta ,\ell )$ are given. We do not need this
level of precision here and will show that (in our setting) the constants $%
L^{\ast }$ and $\delta _{\ast }$ can be chosen arbitrarily large and small,
respectively. Actually, $L^{\ast }(\Xi ,\delta ,\ell )$ tends to infinity as
$\Xi \nearrow \infty $, $\delta \nearrow \delta _{\ast }$ or $\ell \nearrow
\mathcal{N}_{0}^{-1}$. The values of $\delta _{\ast }$ and $L^{\ast }(\Xi
,\delta ,\ell )$ depend only on the geometry of the interaction (that is,
the dimension $d$ and interaction radius $R$ only) and are the same for all
control functions $h:\mathcal{Q}\rightarrow \mathbb{R}_{+}.$
\end{remark}

\subsection{Proof of the uniqueness\label{proof-uniq}}

In this section, we establish the uniqueness of tempered Gibbs measures due
to small activity parameter $z>0$ as stated in Theorem \ref{uniqueness}. For
this, we will use the lattice representation of our model constructed in the
previous section and verify for it both conditions (DP-1) and (DP-2) of
Theorem \ref{Thm_DPuniqueness}.$ $

\textbf{Proof of Theorem \ref{uniqueness}.} According to Proposition \ref%
{lattice2} it is sufficient to prove that $N\left( \mathcal{M}_{h_{F}}(%
\mathfrak{P})\right) \leq 1$. To do so, we check conditions of Theorem \ref%
{Thm_DPuniqueness} for $h:=h_{F}$ defined in Lemma \ref{lattice1}.

A simple change of variables shows that
\begin{equation*}
\int_{\mathcal{Q}}h(\beta )~\mathfrak{m}_{k}^{\overline{\alpha }}(\mathrm{d}%
\beta )=\int_{\widehat{\Gamma }_{k}}F(\widehat{\gamma }_{k})~\mu _{k}^{%
\mathbb{T}^{-1}\overline{\alpha }}(\mathrm{d}\widehat{\gamma }_{k})
\end{equation*}%
for any $\bar{\alpha}\in \mathcal{A}$. Set $\widehat{\eta }:=\mathbb{T}^{-1}%
\overline{\alpha }\in \widehat{\Gamma }(X)$ and observe that $F(\widehat{%
\eta }_{j})=h_{F}(\alpha _{j})$. Corollary \ref{cor2} implies that the
inequality
\begin{equation}
\int_{\widehat{\Gamma }_{k}}F(\widehat{\gamma }_{k})~\mu _{k}^{\widehat{\eta
}}(\mathrm{d}\widehat{\gamma }_{k})\leq \Xi+\delta \sum_{j\in
\partial k}F(\widehat{\eta }_{j}) \label{NEW}
\end{equation}%
holds for any $\delta >0$ with a positive constant $\Xi:=\Xi_{\delta }(\mathcal{J},z),$ which is non-decreasing both in $\mathcal{J}$
and $z$. Thus \textbf{(}DP-1\textbf{)} is proved.$ $

Let us now check \textbf{(}DP-2\textbf{)}. Fix $L>0$ and let $\overline{%
\alpha },\overline{\alpha }^{\prime }\in \cal A$ be boundary conditions satisfying
\begin{equation}
\sup_{j\in \mathbb{Z}^{d}}~\{h(\overline{\alpha }_{j});~h(\overline{\alpha }%
_{j}^{\prime })\}\leq L.  \label{NEW1}
\end{equation}
By a change of variables it is easy to see that
\begin{equation*}
\mathrm{d}_{\mathrm{var}}\left( \mathfrak{m}_{k}^{\overline{\alpha }},~%
\mathfrak{m}_{k}^{\overline{\alpha }^{\prime }}\right) =\mathrm{d}_{\mathrm{%
var}}\left( \mu _{k}^{\widehat{\eta }},~\mu _{k}^{\widehat{\varsigma }%
}\right) \text{ \ \ for }\widehat{\eta }:=\mathbb{T}^{-1}\overline{\alpha },%
\text{ }\widehat{\varsigma }:=\mathbb{T}^{-1}\overline{\alpha }^{\prime }.
\end{equation*}%
Condition (\ref{NEW1}) implies that $\sup_{j}\left\{ F(\widehat{\eta }%
_{j});~F(\widehat{\varsigma }_{j})\right\} =\sup_{j}\{h(\overline{\alpha }%
_{j});~h(\overline{\alpha }_{j}^{\prime })\}\leq L$. Thus, for given $\Xi$, $\delta$ as in \textbf{(}DP-1\textbf{)} and arbitrary $\ell $ and $\mathcal{J}_{0}$, by Lemma \ref{DP2} we can find $z_{0}>0$ such that the bound
$$\mathrm{d}_{\mathrm{var}}(\mu _{k}^{\widehat{\eta }},~\mu _{k}^{\widehat{\varsigma }})\leq \ell $$
 holds uniformly for any $z\leq z_{0}$, $%
\mathcal{J\leq J}_{0}$ and all $\widehat{\eta },\widehat{\varsigma }$ such
that $F(\widehat{\eta }_{j}),F(\widehat{\varsigma }_{j})\leq L$. This completes the proof. \hfill $\square $

\section{Proofs of auxiliary results\label{sec-proofs}}

Our first aim is to prove Proposition \ref{prop_main_est}\textbf{. }We start
with some preparations.

\begin{lemma}
\label{lemma01}For any $\gamma ,\eta \in \Gamma (X)$ and $k\in \mathbb{Z}%
^{d} $ we have the estimate
\begin{equation}
-U_{k}(\gamma _{k}|\eta )\leq -A_{\Phi }N(\gamma _{k})^{P}+\frac{M\mathcal{N}%
_{0}}{2}N(\gamma _{k})^{2}+B_{\Phi }N(\gamma _{k})+\frac{M}{2}\sum_{j\in
\partial k}N(\eta _{j})^{2}.  \label{U-est}
\end{equation}
\end{lemma}

\textbf{Proof. }By definition (\ref{U1}) of the conditional energy $%
U_{k}(\gamma _{k}|\eta )$ and assumptions (A1)--(A3) on $\Phi (x,y)$, we
immediately obtain%
\begin{align}
-U_{k}(\gamma _{k}|\eta )& =-U(\gamma _{k})-\sum_{x\in \gamma
_{k}}\sum_{y\in \eta _{\partial k}}\Phi (x,y)  \label{UQ1} \\
& \leq -\left[ A_{\Phi }N(\gamma _{k})^{P}-B_{\Phi }N(\gamma _{k})\right]
+MN(\gamma _{k})\sum_{j\in \partial k}N(\eta _{j})  \notag \\
& =-A_{\Phi }N(\gamma _{k})^{P}+\frac{M\mathcal{N}_{0}}{2}N(\gamma
_{k})^{2}+B_{\Phi }N(\gamma _{k})+\frac{M}{2}\sum_{j\in \partial k}N(\eta
_{j})^{2},  \notag
\end{align}%
and the proof is complete.
\hfill%
$\square $

\begin{lemma}
\label{lemma1}For any $\varepsilon >0$ the spin-spin energy $E_{k}(\sigma
_{k}|\xi )$ satisfies the following estimate:%
\begin{multline}
-\mathcal{J}^{-1}E_{k}(\sigma _{k}|\xi )\leq \Big[(\mathcal{N}%
_{0}+1)\sum_{x\in \gamma _{k}}\left\vert \sigma _{x}\right\vert
^{r(1+\varepsilon )}+\sum_{j\in \partial k}\sum_{y\in \eta _{j}}\left\vert
\xi _{y}\right\vert ^{r(1+\varepsilon )}\Big]  \label{EL} \\
+\left( 1+\frac{1}{2}C_{W}\right) \Big[(\mathcal{N}_{0}+1)N(\gamma
_{k})^{2+\varepsilon ^{-1}}+\sum_{j\in \partial k}N(\eta
_{j})^{2+\varepsilon ^{-1}}\Big]
\end{multline}%
for all $k\in \mathbb{Z}^{d}$ and $\sigma _{k}\in S^{\gamma _{k}},\xi \in
S^{\eta }$.
\end{lemma}

\textbf{Proof.} By definition (\ref{E1}) of $E_k(\sigma
_{k}|\xi )$ we have%
\begin{equation}
-E_k(\sigma _{k}|\xi )\leq \sum_{\left\{ x,y\right\}
\subset \gamma _{k}}W_{xy}^{-}(\sigma _{x},\sigma _{y})+\sum_{x\in \gamma
_{k}}\sum_{y\in \eta _{\partial k}}W_{xy}^{-}(\sigma _{x},\xi _{y}).
\label{en-est1a}
\end{equation}%
Let us estimate each sum in (\ref{en-est1a}) by means of the classical Young
inequality%
\begin{equation}
ab\leq \frac{a^{\mathrm{p}}}{\mathrm{p}}+\frac{b^{\mathrm{q}}}{\mathrm{q}},%
\text{ for }a,b\geq 0\text{ and }\mathrm{p},\mathrm{q}>1\text{ s.\thinspace
t. }\mathrm{p}^{-1}+\mathrm{q}^{-1}=1.  \label{young}
\end{equation}%
To this end, observe that $\frac{1}{1+\varepsilon }+\frac{1}{1+\varepsilon
^{-1}}=1$ for any $\varepsilon >0$. Using (A4) and then (\ref{young}), we
get
\begin{multline}
\mathcal{J}^{-1}\sum_{\left\{ x,y\right\} \subset \gamma
_{k}}W_{xy}^{-}(\sigma _{x},\sigma _{y})\leq \sum_{\left\{ x,y\right\}
\subset \gamma _{k}}\left( \left\vert \sigma _{x}\right\vert ^{r}+\left\vert
\sigma _{y}\right\vert ^{r}+C_{W}\right)  \label{y1} \\
\leq \left[ N(\gamma _{k})-1\right] \sum_{x\in \gamma _{k}}\left\vert \sigma
_{x}\right\vert ^{r}+C_{W}\frac{N(\gamma _{k})\left[ N(\gamma _{k})-1\right]
}{2} \\
\leq \sum_{x\in \gamma _{k}}\left[ \frac{\left\vert \sigma _{x}\right\vert
^{r(1+\varepsilon )}}{1+\varepsilon }+\frac{N(\gamma _{k})^{1+\varepsilon
^{-1}}}{1+\varepsilon ^{-1}}\right] +\frac{1}{2}C_{W}N(\gamma _{k})^{2} \\
\leq \sum_{x\in \gamma _{k}}\left\vert \sigma _{x}\right\vert
^{r(1+\varepsilon )}+\left( 1+\frac{1}{2}C_{W}\right) N(\gamma
_{k})^{2+\varepsilon ^{-1}}.
\end{multline}%
Similarly, for each $j\in \partial k$ we have
\begin{multline}
\mathcal{J}^{-1}\sum_{x\in \gamma _{k}}\sum_{y\in \eta
_{j}}W_{xy}^{-}(\sigma _{x},\xi _{y})\leq \sum_{x\in \gamma _{k}}\sum_{y\in
\eta _{j}}\left( \left\vert \sigma _{x}\right\vert ^{r}+\left\vert \xi
_{y}\right\vert ^{r}+C_{W}\right)  \label{y2} \\
\leq N(\eta _{j})\sum_{x\in \gamma _{k}}\left\vert \sigma _{x}\right\vert
^{r}+N(\gamma _{k})\sum_{y\in \eta _{j}}\left\vert \xi _{y}\right\vert
^{r}+C_{W}N(\gamma _{k})N(\eta _{j}) \\
\leq \sum_{x\in \gamma _{k}}\left[ \left\vert \sigma _{x}\right\vert
^{r(1+\varepsilon )}+N(\eta _{j})^{1+\varepsilon ^{-1}}\right] +\sum_{y\in
\eta _{j}}\left[ \left\vert \xi _{y}\right\vert ^{r(1+\varepsilon
)}+N(\gamma _{k})^{1+\varepsilon ^{-1}}\right] \\
+\frac{1}{2}C_{W}\left[ N(\gamma _{k})^{2}+N(\eta _{j})^{2}\right] .
\end{multline}%
Another application of Young's inequality yields the bound
\begin{equation*}
N(\gamma _{k})N(\eta _{j})^{1+\varepsilon ^{-1}}\leq N(\gamma
_{k})^{2+\varepsilon ^{-1}}\frac{1}{2+\varepsilon ^{-1}}+N(\eta
_{j})^{2+\varepsilon ^{-1}}\frac{1+\varepsilon ^{-1}}{2+\varepsilon ^{-1}},
\end{equation*}%
by which we conclude that%
\begin{multline}
\mathrm{LHS}(\ref{y2})\leq \sum_{x\in \gamma _{k}}\left\vert \sigma
_{x}\right\vert ^{r(1+\varepsilon )}+\sum_{y\in \eta _{j}}\left\vert \xi
_{y}\right\vert ^{r(1+\varepsilon )}  \label{y4} \\
+\left( 1+\frac{1}{2}C_{W}\right) \left[ N(\gamma _{k})^{2+\varepsilon
^{-1}}+N(\eta _{j})^{2+\varepsilon ^{-1}}\right] .
\end{multline}%
Combining (\ref{y1})--(\ref{y4}), we obtain the desired estimate (\ref{EL}).
\hfill%
$\square \medskip $

\begin{lemma}
\label{superstab} For any $\epsilon >0$ there exists a constant $D_{\epsilon
}>0$ such that the following superstability bound holds:%
\begin{multline}
-H_{k}(\widehat{\gamma }_{k}|\widehat{\eta })+A_{\Phi }N(\gamma _{k})^{P}
\label{y6a} \\
\leq \epsilon \left[ N(\gamma _{k})^{P}+\sum_{x\in \gamma _{k}}\left\vert
\sigma _{x}\right\vert ^{q}+\sum_{j\in \partial k}\left( N(\eta
_{j})^{P}+\sum_{y\in \eta _{j}}\left\vert \xi _{y}\right\vert ^{q}\right) %
\right] +D_{\epsilon },
\end{multline}%
for all $\widehat{\gamma },\widehat{\eta }\in \widehat{\Gamma }(X)$ and $%
k\in \mathbb{Z}^{d}$. Furthermore, $D_{\epsilon }:=D_{\epsilon }(\mathcal{J}%
) $ can be chosen as a non-decreasing functions of $\mathcal{J}$.
\end{lemma}

\textbf{Proof.} It readily follows from (\ref{U-est}) and (\ref{EL}) that
\begin{align}
-H_{k}(\widehat{\gamma }_{k}|\widehat{\eta })& \leq -A_{\Phi }N(\gamma
_{k})^{P}+B_{\Phi ,\mathcal{J}}N(\gamma _{k})^{2+\varepsilon ^{-1}}+C_{%
\mathcal{J}}\sum_{j\in \partial k}N(\eta _{j})^{2+\varepsilon ^{-1}}
\label{H-bound0} \\
& +\mathcal{J}\Big[(\mathcal{N}_{0}+1)\sum_{x\in \gamma _{k}}\left\vert
\sigma _{x}\right\vert ^{r(1+\varepsilon )}+\sum_{y\in \eta _{\partial
k}}\left\vert \xi _{y}\right\vert ^{r(1+\varepsilon )}\Big]  \notag
\end{align}%
for any $\widehat{\gamma },\widehat{\eta }\in \widehat{\Gamma }(X)$, $k\in
\mathbb{Z}^{d}$ and $\varepsilon >0$. Here
\begin{equation}
B_{\Phi ,\mathcal{J}}:=B_{\Phi }+C_{\mathcal{J}}(\mathcal{N}_{0}+1),\text{ \
\ }C_{\mathcal{J}}:=\frac{M}{2}+\mathcal{J}\left( 1+\frac{1}{2}C_{W}\right) ,
\label{H1}
\end{equation}%
are both non-decreasing functions of $\mathcal{J}$. Now let us fix some $%
\varepsilon >0$ such that
\begin{equation}
t:=r(1+\varepsilon )<q\text{ \ and \ }p:=2+\varepsilon ^{-1}<P,  \label{tp}
\end{equation}%
which is possible due to assumption (A6). Note that by (\ref{young}) we have
for any $\theta _{1},\theta _{2}>0$%
\begin{eqnarray}
\sum_{x\in \gamma _{k}}\left\vert \sigma _{x}\right\vert ^{t} &\leq &\theta
_{1}\sum_{x\in \gamma _{k}}\left\vert \sigma _{x}\right\vert ^{q}+\theta
_{1}^{\frac{t}{t-q}}N(\gamma _{k}),  \label{tp1} \\
N(\gamma _{k})^{p} &\leq &\theta _{2}N(\gamma _{k})^{P}+\theta _{2}^{\frac{p%
}{p-P}}.  \label{tp2}
\end{eqnarray}%
Substituting both (\ref{tp1}) and (\ref{tp2}) into (\ref{H-bound0}) and then
taking $\theta _{1},\theta _{2}$ small enough we get the required result.
\hfill%
$\square  $

\noindent \textbf{Proof of Proposition \ref{prop_main_est} }. For any given $\delta$ the estimate (\ref{bound1})
follows immediately from Lemma \ref{superstab} with $\epsilon =\delta \max
\{1,\kappa ,\vartheta \}$ and $C_{\delta }(\kappa ,\vartheta ,\mathcal{J}%
)=D_{\epsilon }(\mathcal{J})$.
\hfill%
$\square $

\begin{remark}
\label{cor3} For $\widehat{\eta }=\emptyset $ we have the (slightly stronger
than (\ref{H-bound0}) and (\ref{H1})) bound
\begin{equation}
-H(\widehat{\gamma }_{k})\leq -A_{\Phi }N(\gamma _{k})^{P}+B_{\Phi ,\mathcal{%
J}}^{0}N(\gamma _{k})^{2+\varepsilon ^{-1}}+\mathcal{J}\sum_{x\in \gamma
_{k}}\left\vert \sigma _{x}\right\vert ^{r(1+\varepsilon )}  \label{H2}
\end{equation}%
where the constant $B_{\Phi ,\mathcal{J}}^{0}:=B_{\Phi }+\mathcal{J}\left( 1+%
\frac{1}{2}C_{W}\right) $ is independent of $\varepsilon >0. $
\end{remark}

\textbf{Proof of Lemma \ref{DP2}. }To keep track of the dependence on the
model parameters ($z$ and $\mathcal{J}$ in particular), all constants in the
estimates below will be written explicitly (although they need not be the
best possible).

The general formula for the total variation distance between two probability
measures states that
\begin{multline}
\mathrm{d}_{\mathrm{var}}(\mu _{k}(\mathrm{d}\widehat{\gamma }_{k}|\widehat{%
\eta }),~\mu _{k}(\mathrm{d}\widehat{\gamma }_{k}|\widehat{\varsigma }))
\label{y0} \\
=\frac{1}{2}\int_{\widehat{\Gamma }_{k}}\left\vert Z_{k}^{-1}(\widehat{\eta }%
)\exp \left\{ -H_{k}(\widehat{\gamma }_{k}|\widehat{\eta })\right\}
-Z_{k}^{-1}(\widehat{\varsigma })\exp \{-H_{k}(\widehat{\gamma }_{k}|%
\widehat{\varsigma })\}\right\vert ~\widehat{\lambda }_z(\mathrm{d}\widehat{%
\gamma }_{k}).
\end{multline}%
Multiplying the right-hand side by the expression $Z_{k}(\widehat{\eta }%
)Z_{k}(\widehat{\varsigma })\geq 1$ and using (\ref{norm_fact}), we see by
an elementary calculation that%
\begin{multline}
\mathrm{d}_{\mathrm{var}}(\mu _{k}(\mathrm{d}\widehat{\gamma }_{k}|\widehat{%
\eta }),~\mu _{k}(\mathrm{d}\widehat{\gamma }_{k}|\widehat{\varsigma }))\leq
\min \left\{ Z_{k}(\widehat{\eta }),\text{ }Z_{k}(\widehat{\varsigma }%
)\right\}  \label{dvar} \\
\times \int_{\widehat{\Gamma }_{k}}\left\vert \exp \{-H_{k}(\widehat{\gamma }%
_{k}|\widehat{\eta })\}-\exp \{-H_{k}(\widehat{\gamma }_{k}|\widehat{%
\varsigma })\}\right\vert ~\widehat{\lambda }_z(\mathrm{d}\widehat{\gamma }%
_{k}).
\end{multline}%

For simplicity, let us first set $\widehat{\varsigma }=\emptyset $ so that $%
H_{k}(\widehat{\gamma }_{k}|\emptyset )=H_{k}(\widehat{\gamma }_{k})$.
Observe that $H_{k}(\widehat{\gamma }_{k}|\widehat{\eta })=H_{k}(\widehat{%
\gamma }_{k})=0$ for $\widehat{\gamma }_{k}=\emptyset $. Therefore
\begin{multline}
\int_{\widehat{\Gamma }_{k}}\left\vert \exp \{-H_{k}(\widehat{\gamma }_{k}|%
\widehat{\eta })\}-\exp \{-H_{k}(\widehat{\gamma }_{k})\}\right\vert ~%
\widehat{\lambda }_z(\mathrm{d}\widehat{\gamma }_{k})  \label{y3} \\
=\int_{\widehat{\Gamma }_{k}\setminus \{\emptyset \}}\left\vert 1-\exp
\left\{ -\mathrm{\Delta }H_{k}(\widehat{\gamma }_{k}|\widehat{\eta }%
)\right\} \right\vert \exp \left\{ -H_{k}(\widehat{\gamma }_{k})\right\} \
\widehat{\lambda }_z(\mathrm{d}\widehat{\gamma }_{k}),
\end{multline}%
where, cf. (\ref{DH}),%
\begin{equation*}
\mathrm{\Delta }H_{k}(\widehat{\gamma }_{k}|\widehat{\eta }):=\sum_{x\in
\gamma _{k},y\in \eta _{\partial k}}\left[ \Phi (x,y)+W_{xy}(\sigma _{x},\xi
_{y})\right] .
\end{equation*}%
Obviously,%
\begin{align}
& \max \left\{ \exp _{{}}^{{}}\left[ -\mathrm{\Delta }H_{k}(\widehat{\gamma }%
_{k}|\widehat{\eta })\right] ,\text{\ }\left\vert 1-\exp _{{}}^{{}}\left[ -%
\mathrm{\Delta }H_{k}(\widehat{\gamma }_{k}|\widehat{\eta })\right]
\right\vert \right\}  \label{y11} \\
& \leq \exp \left\{ [\mathrm{\Delta }H_{k}(\widehat{\gamma }_{k}|\widehat{%
\eta })]^{-}\right\} \leq \exp \left\{ \sum_{x\in \gamma _{k}}\sum_{y\in
\eta _{\partial k}}\left[ \Phi ^{-}(x,y)+W_{xy}^{-}(\sigma _{x},\xi _{y})%
\right] \right\} ,  \notag
\end{align}%
where superscript $^{-}$ denote the negative part of the corresponding
function.\emph{ }

Recall that $\widehat{\eta }=(\eta ,\xi )\in \widehat{\Gamma }(X)$ has to
obey the bound $\sup_{j}F(\widehat{\eta }_{j})\leq L$. Hence
\begin{equation}
\sup_{j\in \mathbb{Z}^{d}}\left\{ N(\eta _{j})^{p},\text{ }\sum_{y\in \eta
_{j}}\left\vert \xi _{y}\right\vert ^{q}\right\} \leq \frac{L}{\max \{\kappa
,\vartheta \}}=:\mathcal{L}  \label{DP_bounds}
\end{equation}
for all $1\leq p\leq P$. Moreover, by (\ref{young}) a similar estimate also
holds for any $1\leq r\leq q$:
\begin{equation}
\sum_{y\in \eta _{j}}\left\vert \xi _{y}\right\vert ^{r}\leq \frac{r}{q}%
\sum_{y\in \eta _{j}}\left\vert \xi _{y}\right\vert ^{q}+\frac{q-r}{q}N(\eta
_{j})\leq \mathcal{L}.  \label{DP3}
\end{equation}%
\emph{ }

Temporarily writing $N$ for $N(\gamma _{k})$ and taking into account that $%
\Phi ^{-}\leq M$, we immediately see by (\ref{UQ1}) and (\ref{DP_bounds})
that%
\begin{equation}
\sum_{x\in \gamma _{k}}\sum_{y\in \eta _{\partial k}}\Phi ^{-}(x,y)\leq MN%
\mathcal{N}_{0}\mathcal{L}.  \label{ya}
\end{equation}%
Next, we fix $\varepsilon >0$, $t\in (r,q)$ and $p\in (2,P)$ as in (\ref{tp}%
). Then, by (\ref{y2}) and (\ref{DP3}) we have%
\begin{multline}
\mathcal{J}^{-1}\sum_{x\in \gamma _{k}}\sum_{y\in \eta _{\partial
k}}W_{xy}^{-}(\sigma _{x},\xi _{y})\leq \mathcal{N}_{0}\mathcal{L}\sum_{x\in
\gamma _{k}}\left\vert \sigma _{x}\right\vert ^{r}+N\sum_{y\in \eta
_{\partial k}}\left\vert \xi _{y}\right\vert ^{r}+C_{W}N\mathcal{N}_{0}%
\mathcal{L}  \label{y8b} \\
\leq \mathcal{N}_{0}\mathcal{L}\left[ \sum_{x\in \gamma _{k}}\left\vert
\sigma _{x}\right\vert ^{r}+\left( 1+C_{W}\right) N\right] .
\end{multline}%
Combining the above inequalities with the superstability bound (\ref{H2}) on
$H_{k}(\widehat{\gamma }_{k})$ and then setting
\begin{equation*}
B_{\Phi ,\mathcal{J}}^{\prime }:=B_{\Phi }+\mathcal{N}_{0}\mathcal{L}M+%
\mathcal{J}\left( 1+C_{W}\right) \left( 1+\mathcal{N}_{0}\mathcal{L}\right) ,
\end{equation*}%
we obtain the estimate%
\begin{multline}
\max \left\{ -H_{k}(\widehat{\gamma }_{k}|\widehat{\eta }),\text{\ }-H_{k}(%
\widehat{\gamma }_{k})+\ln \left\vert 1-\exp \left\{ -\mathrm{\Delta }H_{k}(%
\widehat{\gamma }_{k}|\widehat{\eta })\right\} \right\vert \right\}
\label{y3a} \\
\leq -A_{\Phi }N^{P}+B_{\Phi ,\mathcal{J}}^{\prime }N^{p}+\mathcal{J}%
\sum_{x\in \gamma _{k}}\left[ \left\vert \sigma _{x}\right\vert ^{t}+%
\mathcal{N}_{0}\mathcal{L}\left\vert \sigma _{x}\right\vert ^{r}\right] .
\end{multline}%
Notice that by Young's inequality the following uniform bound holds:
\begin{equation}
C_{\Phi ,\mathcal{J}}:=\max_{N\geq 0}\left\{ -A_{\Phi }N^{P}+B_{\Phi ,%
\mathcal{J}}^{\prime }N^{p}\right\} \leq \left( A_{\Phi }\right) ^{-\frac{p}{%
P-p}}\left( B_{\Phi ,\mathcal{J}}^{\prime }\right) ^{\frac{P}{P-p}}.
\label{y5}
\end{equation}%
Thereafter, using the disintegration (\ref{LPM}) we conclude (analogously to
(\ref{finexp})) that
\begin{multline}
\mathrm{RHS\ }\text{(\ref{y3})}  \label{y7} \\
\leq e^{C_{\Phi ,\mathcal{J}}}\int_{\Gamma _{k}\setminus \{\emptyset
\}}\int_{S^{\gamma _{k}}}\exp \left\{ \mathcal{J}\sum_{x\in \gamma _{k}}%
\left[ \left\vert \sigma _{x}\right\vert ^{t}+\mathcal{N}_{0}\mathcal{L}%
\left\vert \sigma _{x}\right\vert ^{r}\right] \right\} ~\bigotimes_{x\in
\gamma _{k}}\chi (\mathrm{d}\sigma _{x})\lambda_z (\mathrm{d}\gamma _{k}) \\
=e^{C_{\Phi ,\mathcal{J}}}\int_{\Gamma _{k}\setminus \{\emptyset \}}\left[
\mathcal{E}_{_{\mathcal{J}}}\right] ^{N(\gamma _{k})}~\lambda_z (\mathrm{d}%
\gamma _{k})=e^{C_{\Phi ,\mathcal{J}}}\sum_{n=1}^{\infty }\frac{\left( z%
\mathcal{E}_{_{\mathcal{J}}}\right) ^{n}}{n!}=e^{C_{\Phi ,\mathcal{J}}}\left[
\exp \left\{ z\mathcal{E}_{_{\mathcal{J}}}\right\} -1\right] ,
\end{multline}%
where
\begin{equation}
\mathcal{E}_{_{\mathcal{J}}}:=\int_{S}\exp \left\{ \mathcal{J}\left(
\left\vert s\right\vert ^{t}+\mathcal{N}_{0}\mathcal{L}\left\vert
s\right\vert ^{r}\right) \right\} ~\chi (\mathrm{d}s)  \label{y9}
\end{equation}%
is finite by assumption (A5).\emph{ }

We proceed in a similar way to obtain an upper bound on $Z_{k}(\widehat{\eta
})$. Indeed, with the help of (\ref{y3a})--(\ref{y9}) one gets%
\begin{multline}
Z_{k}(\widehat{\eta }):=\int_{\widehat{\Gamma }_{k}}\exp \{-H_{k}(\widehat{%
\gamma }_{k}|\widehat{\eta })\}~\widehat{\lambda }_z(\mathrm{d}\widehat{\gamma
}_{k})  \label{y10a} \\
\leq e^{C_{\Phi ,\mathcal{J}}}\int_{\Gamma _{k}}\int_{S^{\gamma _{k}}}\exp
\left\{ \mathcal{J}\sum_{x\in \gamma _{k}}\left( \left\vert \sigma
_{x}\right\vert ^{t}+\mathcal{N}_{0}\mathcal{L}\left\vert \sigma
_{x}\right\vert ^{r}\right) \right\} ~\bigotimes_{x\in \gamma _{k}}\chi (%
\mathrm{d}\sigma _{x})\lambda_z (\mathrm{d}\gamma _{k}) \\
=\exp \left\{ C_{\Phi ,\mathcal{J}}+z\mathcal{E}_{_{\mathcal{J}}}\right\} .
\end{multline}%
Putting (\ref{y7}) and (\ref{y10a}) together and using the well-known inequality $e^a-1\le ae^a$ for all $a\ge 0$,
we conclude that%
\begin{equation}
\mathrm{d}_{\mathrm{var}}(\mu _{k}(\mathrm{d}\widehat{\gamma }_{k}|\widehat{%
\eta }),\text{ }\mu _{k}(\mathrm{d}\widehat{\gamma }_{k}|\emptyset ))\leq z%
\mathcal{E}_{_{\mathcal{J}}}\exp \left\{ 2\left( C_{\Phi ,\mathcal{J}}+z%
\mathcal{E}_{_{\mathcal{J}}}\right) \right\} .  \label{dest2}
\end{equation}%
By the triangle inequality the above bound extends to general boundary
conditions $\widehat{\varsigma }\neq \emptyset $. This yields the desired
estimate (\ref{var1}) with
\begin{equation*}
\phi (z,\mathcal{J},L):=2\mathcal{E}_{_{\mathcal{J}}}\exp \left\{ 2\left(
C_{\Phi ,\mathcal{J}}+z\mathcal{E}_{_{\mathcal{J}}}\right) \right\} ,
\end{equation*}%
which is a non-decreasing function of $\mathcal{J}$, $z$, and $L$.
\hfill%
$\square $

\end{document}